\pdfoutput=1
\documentclass{article}

\usepackage{array}
\usepackage[utf8]{inputenc}
\usepackage{listings}
\usepackage{fancyhdr}
\usepackage{lastpage}
\usepackage{float}
\usepackage{url}
\usepackage{xcolor}
\usepackage{hyperref}
\usepackage{nameref}
\usepackage{enumitem}
\usepackage{adjustbox}
\usepackage{amssymb}
\usepackage{amsthm}
\usepackage{amsmath}
\usepackage{mathtools}
\usepackage{algorithm}
\usepackage{algpseudocode}
\usepackage{algcompatible}
\usepackage[numbers,sort&compress]{natbib}
\usepackage{tikz}
\usepackage{pgfplots}
\usepackage{textcomp}
\usepackage{subcaption}
\usepackage{fontawesome5}
\usepackage{tabularx}
\usepackage{multirow}
\usepackage{booktabs}
\usepackage{longtable}
\usepackage{lscape}
\usepackage[none]{hyphenat}

\graphicspath{{figures/}}

\hyphenchar\font=-1 
\usepackage[margin=1.2truein]{geometry} 
\makeatletter
\renewcommand{\@seccntformat}[1]{}
\makeatother

\hypersetup{
    colorlinks=true,
    linkcolor=red,
    urlcolor=red,
    citecolor=red
}

\usetikzlibrary{
    shapes,
    arrows,
    arrows.meta,
    positioning,
    backgrounds,
    shapes.multipart,
    calc
}

\tikzset{
    block/.style={draw, fill=white, rectangle, minimum height=48pt, minimum width=48pt},
    round/.style={draw, fill=white, circle},
    sum/.style={draw, fill=white, circle, node distance=2cm},
    input/.style={coordinate},
    output/.style={coordinate},
}

\pgfplotsset{compat=newest}

\definecolor{codegreen}{rgb}{0,0.6,0}
\definecolor{codegray}{rgb}{0.5,0.5,0.5}
\definecolor{codepurple}{rgb}{0.58,0,0.82}
\definecolor{backcolour}{rgb}{0.95,0.95,0.92}

\lstdefinelanguage{python}{
    basicstyle=\ttfamily\footnotesize,
    breaklines=true,                 
    captionpos=b,                    
    tabsize=2
}

\lstdefinelanguage{json}{
    basicstyle=\ttfamily\footnotesize,
    showstringspaces=false,
    keywordstyle=\color{blue},
    stringstyle=\color{black},
    string=[s]{"}{"}
}

\lstdefinestyle{sql}{
    language=SQL,
    basicstyle=\ttfamily\footnotesize,
    keywordstyle=\color{blue}\bfseries,
    identifierstyle=\color{black},
    stringstyle=\color{red},
    commentstyle=\color{green},
    morecomment=[l]{--},
    morecomment=[s]{/*}{*/}
}

\begin{document}
\title{A General Capacity Frontier of Complex Networks}
\author{
    Nicholas Kunz \href{mailto:nhk37@cornell.edu}{nhk37@cornell.edu}\\
    H. Oliver Gao \href{mailto:hg55@cornell.edu}{hg55@cornell.edu}\\~\\
    Systems Engineering, Cornell University}
\date{\today}
\maketitle
\thispagestyle{fancy}

\begin{abstract}
\noindent
    From cells to cities, ecosystems to economies, complex networks exhibit peak intensities that define their empirical limit. Estimating these limits traditionally requires domain-specific knowledge. We demonstrate that a wide range of complex networks share a capacity frontier that constrains these maxima. We formalize this regularity by proposing Structural Capacity Theory for Network Substrate Structured Systems. These systems couple a fixed network substrate with an observable dynamical process. The frontier is learned from a vector representation of the network substrate and modulated log-additively by a vector representation of the dynamical process. Frontiers trained on one set of domains reliably bound maxima of entirely different ones without observing their peak intensities or simulating their dynamics. Independent learning paradigms converge on consistent upper bounds. Frontier performance is stable under perturbation, yet degrades under deliberate falsification, with no performance gain under ablation. These findings support the capacity frontier as a generalizable constraint that describes how network structure bounds dynamical processes across otherwise unrelated systems.
\end{abstract}

\vspace{1em}
\phantomsection
\label{sec:intro}
Maxima are fundamental features of complex networks. Understanding them is important because they often mark where critical thresholds are met that cause material changes in network behavior \citep{ref01, ref02, ref16, ref33, ref34}. For example, a regulatory network that exceeds its threshold of localized molecular activity changes regimes \citep{ref25,ref02}. A transportation network that exceeds its carrying capacity changes behavior primarily by congestion, not by design \citep{ref03,ref04}. Although it is known that network maxima can affect network behavior, understanding \textit{how} traditionally requires domain-specific knowledge \citep{ref02,ref03}. Whether or not network maxima have a common regularity across domains is still an open question \citep{ref13,ref14,ref15}. Addressing it requires a shared basis of comparison.

In the previous examples, network structure formed the basis on which a dynamical process occurred \citep{ref02, ref03, ref04, ref05, ref15, ref33}. Size, connectivity, spectral organization, and other topological characteristics constrain network throughput \citep{ref07,ref08,ref28,ref35}, propagation \citep{ref09,ref10,ref30}, synchronization \citep{ref11}, controllability \citep{ref27}, and spatial transport \citep{ref32}. The same network substrate can have different limits depending on the observational definition, activity attribution, temporal aggregation, and measurement regime \citep{ref06,ref26,ref12}. Conversely, structurally distinct substrates and dynamical processes can exhibit comparable limits under shared measurement conventions and operational definitions \citep{ref17,ref18,ref24,ref15}. In both cases, maxima reflect the network substrate and the dynamical process by which observations are attributed, generated, and measured.

\begin{figure}[H]
\centering
\includegraphics[width=\linewidth]{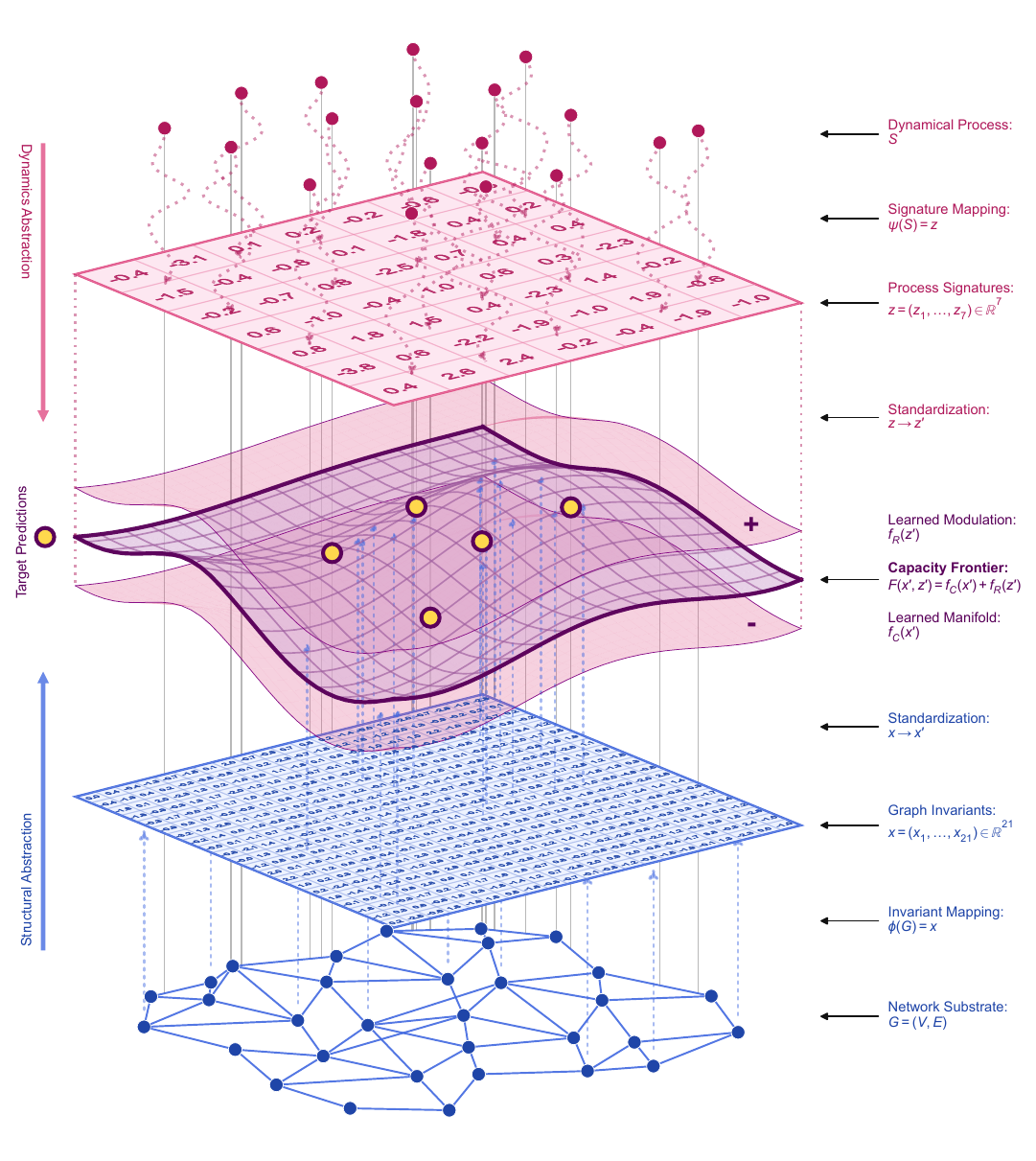}
\caption{
    \textbf{Conceptual representation of Structural Capacity Theory (SCT).} Schematic illustration of Network Substrate Structured Systems (N3S). The network substrate $G$ (blue graph, bottom) is encoded as a standardized vector $x'$ on the structural plane. The dynamical process $S$ (pink markers, top) produces observations attributed to network substrate elements and encoded as a standardized vector $z'$ on the dynamics plane. The curved manifold (purple surface, middle) represents the capacity frontier $F(x', z') = f_C(x') + f_R(z')$, where the learned structural component $f_C(x')$ is modulated log-additively by the learned process component $f_R(z')$ to estimate capacity frontier values $\widehat{y}^*$ (gold points, middle).
}
\label{fig:concept}
\end{figure}
\clearpage

We propose Network Substrate Structured Systems (N3S) as a class of systems for which this coupling can be evaluated. An N3S is defined by a fixed network substrate, an observable dynamical process that corresponds to that substrate, and a measurable maximum rate that corresponds to that dynamical process (Fig.~\ref{fig:concept}). Membership is determined from the observations a given system naturally provides so that the scope of the theory is fixed before it is tested. Structural Capacity Theory (SCT) proposes that maxima from any N3S are comparable when represented in a shared coordinate system, and within that representation, maxima are constrained by a common capacity frontier.

We estimate the capacity frontier as a covariate-dependent upper-envelope \citep{ref103,ref111}. Typical network activity can be captured by central tendencies, but the area of interest is the ceiling, so the estimators are trained as such \citep{ref19,ref20,ref21,ref22,ref23,ref31,ref29}. The system's network substrate $G$ and dynamical process $S$ are encoded as graph invariants $x$ and process signatures $z$, and standardized to $x'$ and $z'$, respectively. The capacity frontier $F(x', z') = f_C(x') + f_R(z')$ (Eq.~\eqref{eq:frontier}) log-additively bounds the log-transformed maximum rate $y^* = \ln(1 + r_{\max})$ (Eq.~\eqref{eq:maximum}). The frontier is strictly empirical. It permits a small fraction of observations to cross it and carries no claim of analytical containment.

We evaluate whether this formulation supports a transferable, consistent, stable, falsifiable, and decomposable capacity frontier. We found a frontier that bounded maxima from out-of-sample systems using Leave-One-Group-Out Cross-Validation (LOGO-CV) and repeated $k$-Fold Cross-Validation ($k$-Fold-CV) with equivalent performance. Independent estimators from 3 distinct learning paradigms predicted consistent frontier values. Perturbations to network substrates, graph invariants, dynamical processes, and process signatures preserved frontier performance. By contrast, 3 falsification procedures that deliberately destroyed the representation–target correspondence degraded frontier performance, which retraining could not recover. Finally, ablations that tested simpler and more complex alternative formulations did not meaningfully improve the performance of the frontier.

\phantomsection
\label{sec:results}

\subsection*{Shared Coordinate System}
\label{sub-sec:shared}
Encoding all systems as paired network substrate and dynamical process vectors placed unrelated phenomena within a shared coordinate system (Fig.~\ref{fig:concept}; Supplementary Table~\ref{tab:si-corpus}). The corpus of 25 systems contained 5 broad domains: Earth \& Physical Sciences, Life Sciences \& Medicine, Technology \& Information, Trade \& Institutions, and Transportation \& Infrastructure. Each of the 5 broad domains contained 5 individual disciplines. Their measurements did not share units before encoding. The network substrate vector $x$ admitted 21 graph invariants (Supplementary Table~\ref{tab:si-features}). The dynamical process vector $z$ admitted 7 process signatures (Supplementary Table~\ref{tab:si-signatures}). Both vectors were selected under prespecified criteria and standardized to $x'$ and $z'$, respectively (\nameref{sec:encoding}). This encoding recast every system with the same number of dimensions despite large differences in substance and scale. It made the diverse corpus of 25 systems directly comparable.

Within this shared representation, the capacity frontier bounded maxima across the corpus (Supplementary Table~\ref{tab:si-transfer}). Under repeated $k$-Fold-CV, the frontier was fit on each training set and evaluated on the corresponding test set (\nameref{sec:resampling}). Each test was repeated 30 times to reduce dependence on any single seeded fold. Frontier performance was reported as medians across estimators after averaging them over test folds and repetitions. Under repeated $5$-Fold-CV and repeated $10$-Fold-CV, frontier efficiency was 0.77 and 0.76, respectively. Frontier efficiency was measured using the Efficiency Index (EI) --- the bounded $[0, 1]$ geometric mean of the non-violation rate and two bounded metrics that decreased with violation magnitude and slack relative to the evaluated target scale (\nameref{sec:performance-metrics},~Eq.~\eqref{eq:ei}). There were infrequent frontier violations, low violation magnitude, and consistent slack. Under repeated 5-Fold-CV, the violation rate was 0.02, mean violation magnitude 0.06, and the mean slack 6.86 log units. Under repeated 10-Fold-CV, the corresponding values were 0.01, 0.02, and 6.87 log units, respectively. 
Thus, the shared representation supported a common capacity frontier.

\subsection*{Transfer Across Domains}
\label{sub-sec:transfer}
\begin{figure}[H]
\centering
\includegraphics[width=\linewidth]{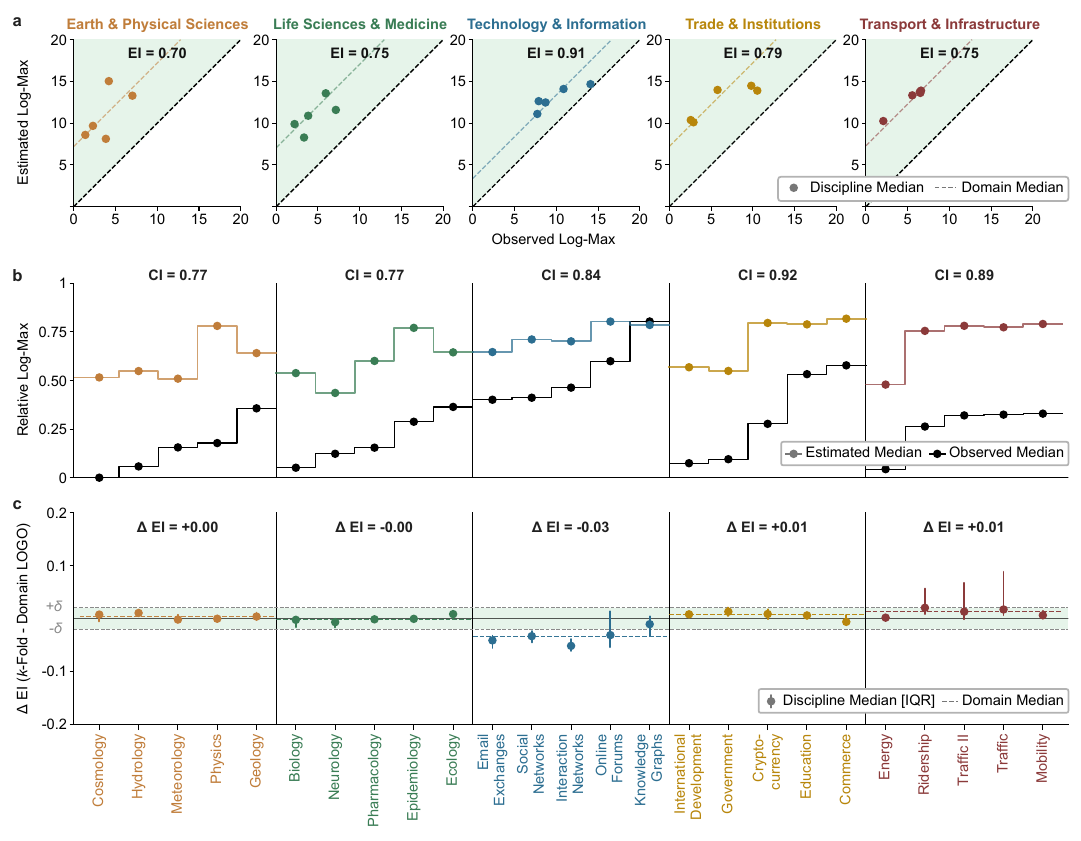}
\caption{
    \textbf{Generalization of the capacity frontier.} Domain-level LOGO-CV across 25 systems in 5 broad domains. \textbf{a}. Estimated versus observed log-maxima for each withheld domain. Discipline medians (domain-color markers) fall along domain offsets (domain-color dashed lines) relative to equality (diagonal gray dashed line). Green shading illustrates estimates at or above log-maxima. \textbf{b}. Estimation-observation consensus profiles within withheld domains. Step traces and markers exhibit observed log-maxima (black trace, bottom) and median model estimates (domain-color trace, top), with disciplines ordered by observed maximum and profiles scaled to the unit interval. \textbf{c}. Paired differences in frontier efficiency between repeated $k$-Fold-CV and domain-level LOGO-CV. Markers with vertical intervals exhibit discipline medians and interquartile ranges across models (domain-color intervals), dashed horizontal lines exhibit domain medians (domain-color dashed lines), and the horizontal shading is the equivalence margin ($\pm\delta_{\mathrm{EI}}$, green shaded band). Annotations report (\textbf{a}) domain EI, (\textbf{b}) domain CI, and (\textbf{c}) domain $\Delta\mathrm{EI}$.
}
\label{fig:transfer}
\end{figure}

A capacity frontier trained without a given domain constrained the maxima of that domain’s disciplines (Supplementary Table~\ref{tab:si-transfer}). Under domain-level LOGO-CV, the frontier was trained on 4 of the 5 broad domains and evaluated on the 5 individual disciplines in the withheld domain. Frontier efficiency was 0.75, ranging from 0.70 to 0.91 (Fig.~\ref{fig:transfer}a). The violation rate was 0.04, mean violation magnitude 0.08, and mean slack 6.86 log units. Estimation-observation consensus was 0.84, ranging from 0.77 to 0.92 across the same evaluations (Fig.~\ref{fig:transfer}b). Estimation-observation consensus was measured using the Consensus Index (CI) --- the bounded $[0, 1]$ geometric mean of global rank agreement, high-capacity overlap, and non-linear dependence (\nameref{sec:performance-metrics},~Eq.~\eqref{eq:ci}). Each withheld discipline supplied its own graph invariants, process signatures, and maximum rate. It tested domain transfer by out-of-sample training, not prediction in the absence of observation. This established generalization across the tested domains.

A capacity frontier trained without a given discipline also bounded that discipline’s maxima. Withholding a single discipline tested generalization similarly to domain-level LOGO-CV, but at a higher resolution. Under discipline-level LOGO-CV the frontier was trained on 24 of the 25 disciplines and evaluated on the one that was withheld, effectively Leave-One-Out Cross-Validation (LOO-CV). In this resampling regime frontier efficiency was 0.73, ranging from 0.52 to 0.90 across disciplines. The violation rate was 0.00, mean violation magnitude 0.00, and mean slack 6.12 log units. Estimation-observation consensus was 0.82 based on out-of-sample predictions pooled within each of the 5 broad domains, ranging from 0.77 to 0.92 across domains. Violations were less common and smaller than under domain-level LOGO-CV. The experiment could not determine whether these differences reflected denser coverage of the coordinate system by in-sample disciplines or differences in test set size and metric aggregation. Nevertheless, the frontier generalized to out-of-sample disciplines similarly to out-of-sample domains.

The capacity frontier achieved equivalent frontier performance across resampling regimes. Using the repeated $k$-Fold-CV results (\nameref{sub-sec:shared}), paired differences in frontier performance between domain-level LOGO-CV and repeated $k$-Fold-CV were tested across disciplines within the 5 broad domains (Fig.~\ref{fig:transfer}c). Paired differences were computed by subtracting domain-level LOGO-CV results from those contained in the comparative resampling regime, where negative values indicated lower performance. Median paired differences in frontier efficiency were 0.00 for all resampling regimes. Median paired differences in estimation-observation consensus were at most 0.01 in magnitude under the same evaluation. Paired Wilcoxon Two One-Sided Tests (TOST) across 45 estimator-by-domain comparisons supported equivalence (\nameref{sec:si-stat-conventions}) within their empirically derived margins ($\delta_{\mathrm{EI}} = 0.02$ and $\delta_{\mathrm{CI}} = 0.11$, all Holm-adjusted $p < 0.001$; Supplementary Table~\ref{tab:si-transfer-equivalence}). Withholding broad domains and individual disciplines preserved frontier performance when compared to repeated random resampling. Therefore, domain generalization did not depend on any single tested resampling regime.

\subsection*{Consensus Across Paradigms}
\label{sub-sec:consensus}
The capacity frontier was consistent across learning paradigms (Fig.~\ref{fig:consensus}). There were 9 independent estimators across 3 distinct learning paradigms, all of which broadly agreed on their predicted frontier values. The learning paradigms were linear parametric, non-linear ensemble, and neural networks (\nameref{sec:frontier-learning}). All models were trained on the same vectors $x'$ and $z'$ while the estimators differed in functional form, regularization, and optimization. Each test was repeated 30 times to reduce sensitivity to parameter initialization and optimization stochasticity. Predicted frontier values were averaged across repetitions. Pairwise model consensus across all 36 pairs of seed-averaged predictions was 0.90 (Fig.~\ref{fig:consensus}a). A geometric mean this close to 1 simultaneously required high rank agreement, top-weighted overlap, and statistical dependence, such that no single component could carry the outcome. Rescaled Spearman rank correlation ($\rho_s^{+}$), Rank-Biased Overlap (RBO), and distance correlation ($\mathrm{dCor}$) were 0.95, 0.86, and 0.91, respectively. The agreement was evaluated on in-sample analyses using full-corpus fits. Consensus across the tested learning paradigms indicated that the frontier was unlikely an artifact of inductive bias (Fig.~\ref{fig:consensus}b).

\subsection*{Stable Under Perturbation}
\label{sub-sec:perturb}
The capacity frontier was stable under perturbation (Fig.~\ref{fig:tests}a, Fig.~\ref{fig:tests}b; Supplementary Tables~\ref{tab:si-perturb} and \ref{tab:si-protocol}). There were 12 perturbation procedures that introduced controlled modifications to the vectors $x'$ and $z'$ while preserving the target $y^*$ across 4 categories: network substrates, graph invariants, dynamical processes, and process signatures (\nameref{sec:stress}). Under domain-level LOGO-CV, the frontier was trained on the original data, the model frozen, then evaluated on the perturbed data in the withheld domain. Each test was repeated 30 times and averaged across repetitions to reduce sensitivity to perturbation realization and estimator initialization. Paired differences were computed by subtracting the original results from the perturbed ones across 45 estimator-by-domain comparisons, where negative values indicated lower performance. At the highest tested intensity, median paired differences in frontier efficiency ranged from -0.02 to +0.01. Median paired differences in estimation-observation consensus were 0.00 for all network and invariant perturbations. Median paired differences in consensus were -0.11 for process bootstrapping, -0.02 for process scaling, and -0.06 for process smoothing. Signature jitter, noise, and subset produced median consensus differences of -0.01, -0.01, and -0.04, respectively. Paired Wilcoxon TOST across 45 estimator-by-domain comparisons supported equivalence for all 12 perturbations within their empirically derived margins ($\delta_{\mathrm{EI}} = 0.08$ and $\delta_{\mathrm{CI}} = 0.14$, all Holm-adjusted $p \leq 0.028$; Supplementary Table~\ref{tab:si-perturb}). Response curves across normalized perturbation intensities were descriptive, where only the highest perturbation intensity was formally tested. Nevertheless, the frontier could withstand structural modifications, process distortions, and feature noise while remaining reliable.

\subsection*{Degradation Under Falsification}
\label{sub-sec:falsify}
The capacity frontier predictably degraded under falsification (Fig.~\ref{fig:tests}c; Supplementary Tables~\ref{tab:si-falsify} and \ref{tab:si-protocol}). There were 3 falsification procedures that deliberately destroyed the correspondence between the vectors $x'$ and $z'$ and the target $y^*$ in distinct ways: target permutation, random generation, and vector generation (\nameref{sec:stress}). Under domain-level LOGO-CV, the frontier was evaluated by freezing the models after training on the original data, and separately after retraining on the falsified data. Each test was repeated 30 times and averaged across repetitions to reduce sensitivity to estimator initialization and training stochasticity. Paired differences were computed by subtracting the original results from the falsified ones across 45 estimator-by-domain comparisons, where negative values indicated lower performance. Under the frozen model, median paired differences in frontier efficiency were -0.02 for target permutation, -0.04 for random generation, and -0.03 for vector generation. After retraining, median paired differences in frontier efficiency were -0.09, -0.02, and -0.03, respectively. Median paired differences in estimation-observation consensus were -0.29, -0.28, and -0.21 when frozen, and -0.24, -0.18, and -0.31 after retraining, respectively. Paired One-Sided Wilcoxon Signed-Rank Tests supported lower efficiency and lower consensus across all 12 comparisons (all Holm-adjusted $p \le 0.013$ for EI and $p < 0.001$ for CI; Supplementary Table~\ref{tab:si-falsify}). The falsified corpora served as negative controls that indicated the frontier depended on a true representation–target correspondence.

\subsection*{Parsimony Under Ablation}
\label{sub-sec:ablate}
\begin{figure}[H]
\centering
\includegraphics[width=\linewidth]{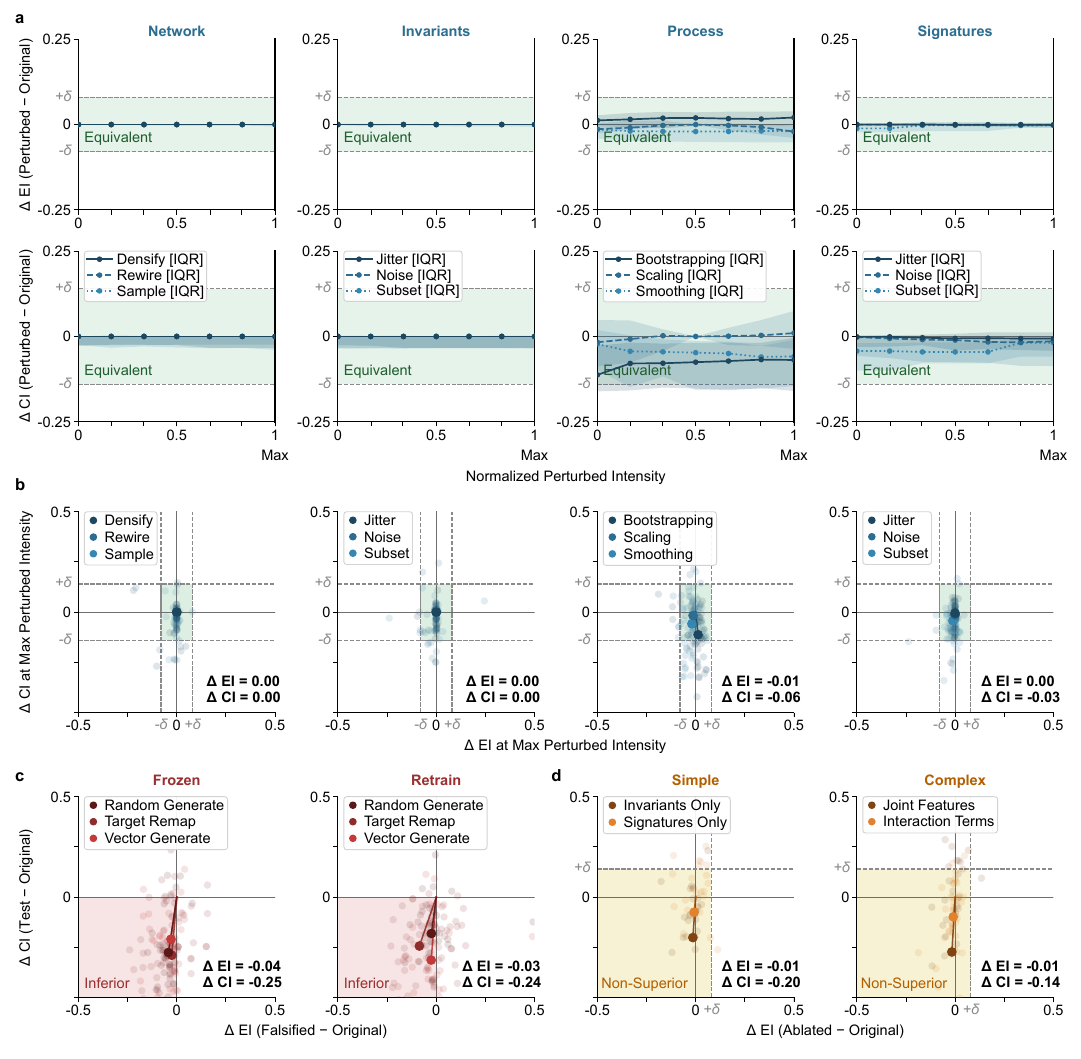}
\caption{
    \textbf{Stress tests of the capacity frontier.} Paired shifts across 45 estimator-by-domain comparisons under domain-level LOGO-CV, computed as test condition minus original condition or alternative formulations minus log-additive decomposition. \textbf{a}. Response curves across normalized intensity for 12 perturbations in 4 categories with $\Delta\mathrm{EI}$ above and $\Delta\mathrm{CI}$ below. Blue lines and shaded bands exhibit medians and IQR, respectively. \textbf{b}. Paired shifts at highest tested perturbation intensity. \textbf{c}. Falsification shifts under frozen and retrain frontiers for 3 falsification procedures. The red quadrant marks joint degradation ($\Delta\mathrm{EI} < 0$, $\Delta\mathrm{CI} < 0$). \textbf{d}. Simple (Invariants Only, Signatures Only) and Complex (Joint Features, Interaction Terms) alternative formulations compared with the log-additive decomposition. In \textbf{b--d}, faint points exhibit individual comparisons and solid points exhibit coordinate-wise medians connected to the origin. Green regions (\textbf{a}, \textbf{b}) mark perturbation equivalence margins ($\delta_{\mathrm{EI}} = 0.08$, $\delta_{\mathrm{CI}} = 0.14$), and the yellow region (\textbf{d}) marks the ablation non-inferiority margin ($\delta_{\mathrm{EI}} = 0.08$, $\delta_{\mathrm{CI}} = 0.14$).
}
\label{fig:tests}
\end{figure}

No alternative capacity frontier formulation meaningfully outperformed the log-additive decomposition (Fig.~\ref{fig:tests}d; Supplementary Tables~\ref{tab:si-ablation} and \ref{tab:si-protocol}). The 4 alternative specifications contained simple ablations: invariants only and signatures only, and complex ablations: joint features and interaction terms (\nameref{sec:stress}). Invariants only trained the frontier as $F(x')$ alone, while signatures only used $F(z')$ alone. Joint features concatenated both vectors into a single-stage fit $F([x', z'])$, while interaction terms augmented the log-additive decomposition as $F(x', z', h')$ with pairwise products $x_j z_k$. Under domain-level LOGO-CV, ablations were fit on each training set and evaluated on the corresponding test set. Each test was repeated 30 times and averaged across repetitions to reduce sensitivity to estimator initialization and training stochasticity. Paired differences were computed by subtracting the original results from the ablated ones across 45 estimator-by-domain comparisons, where negative values indicated lower performance. Median paired differences in frontier efficiency were -0.02 for joint features, and -0.01 for invariants only, signatures only, and interaction terms. Median paired differences in estimation-observation consensus were -0.20 for invariants only, -0.08 for signatures only, -0.27 for joint features, and -0.10 for interaction terms. Paired One-Sided Wilcoxon Signed-Rank tests supported non-inferiority of the log-additive decomposition to all 4 alternatives in both metrics within their empirically derived margins ($\delta_{\mathrm{EI}} = 0.08$ and $\delta_{\mathrm{CI}} = 0.14$, all Holm-adjusted $p < 0.001$; Supplementary Table~\ref{tab:si-ablation}). These tests limited alternatives and supported the log-additive decomposition as the parsimonious formulation of the frontier.

\begin{figure}[H]
\centering
\includegraphics[width=\linewidth]{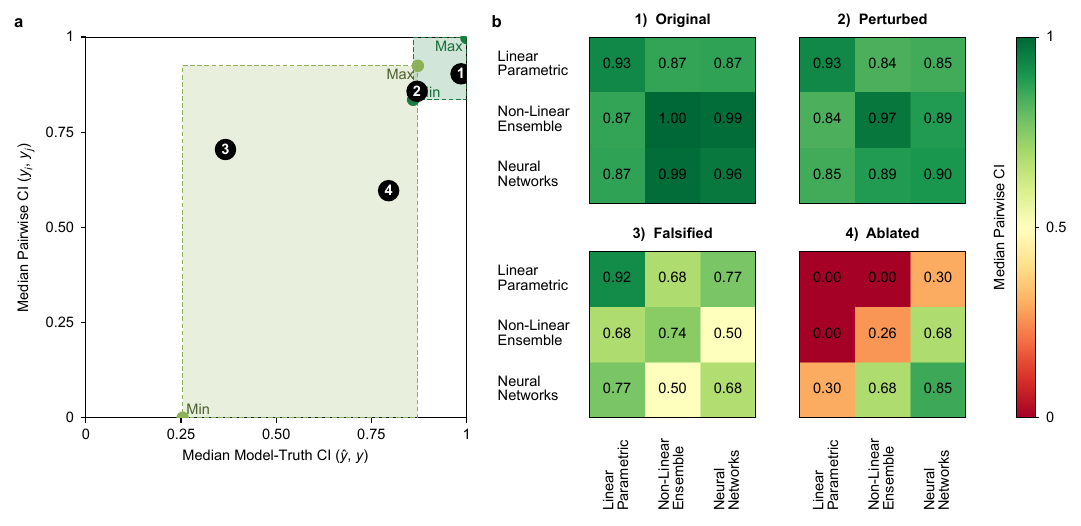}
\caption{
    \textbf{Learning paradigm and estimation-observation consensus.} 9 estimators across 3 distinct learning paradigms agree on full-corpus fits. \textbf{a}. Median estimation-observation consensus ($\widehat{y}^*$ versus $y^*$) by median pairwise model consensus ($\widehat{y}^*_i$ versus $\widehat{y}^*_j$). Numbered points (black markers, white text) label the tests: (1) original, (2) perturbed, (3) falsified, and (4) ablated. Shaded rectangles span coordinate-wise ranges of learning paradigm medians for the original and perturbed conditions (dashed polygon, green) and falsified and ablated conditions (dashed polygon, yellow). \textbf{b}. Median pairwise model consensus across learning paradigms for each test in \textbf{a} with paradigm agreement along the diagonal axes.
}
\label{fig:consensus}
\end{figure}

\section*{Discussion}
\label{sec:discussion}

The capacity frontier constrains maxima across disciplinary boundaries. A frontier with either a broad domain or individual discipline withheld from training bounded maximum rates with performance comparable to repeated random resampling. Information from one system was therefore useful for estimating maxima in another despite differences in subject, size, scale, units, and dynamics. These comparisons provide a basis for asking which constraints are shared across systems and which depend on domain specificity. Capacity estimation can therefore use empirical evidence beyond a system’s own substantive boundary.

Furthermore, the capacity frontier does not require a model of the dynamical processes. Graph invariants and process signatures support capacity estimation without training domain-specific models or simulating their dynamics. Domain-specific models are still important for causal explanation and intervention, but SCT provides a standard benchmark as a complement. Interpreting this benchmark requires additional inquiry, as the reported slack was nearly 3 orders of magnitude higher than the maximum rate. This gap may reflect a loose envelope, unequal observation windows, differences in measurements, or a combination of them. Although the meaning of the slack is still unclear, the frontier still reliably bounds maxima without a traditional model of the underlying dynamics.

Selective stability demonstrates why the capacity frontier is informative. Perturbations at the highest tested intensity generally preserved the frontier's original performance. Falsifications that deliberately destroyed the representation–target correspondence degraded frontier performance, which was unrecoverable even after retraining. Consensus among learning paradigms was relatively high and comparatively similar between the original and perturbed conditions. Only after falsification and ablation did the model consensus between learning paradigms degrade. This supports the frontier as a non-spurious object, distinguishable from a fragile upper-envelope.

The capacity frontier is jointly determined and decomposable. Removing either component or combining them without log-additivity did not improve frontier performance or consensus among estimators. The results support the log-additive decomposition, but leave its exhaustiveness unresolved. All findings were conditioned on the 25 tested systems, 21 graph invariants, 7 process signatures, 9 learned estimators, the log-additive decomposition, and the daily temporal aggregation. Broader corpora, different vector representations, more diverse learning paradigms, and alternative temporal scales all provide direct tests of SCT.

\section*{Methods}
\label{sec:methods}

\subsection*{N3S Definition}
Within SCT, an N3S contains a fixed network substrate $G$, an observable dynamical process $S$, and a measurable maximum rate $r_{\max}$. The network substrate contains finite elements with relational or attribution structure. The dynamical process is a time-indexed sequence of observed events or reconstructed events from published schedules. Each event is attributable or relates to one or more elements of the network substrate. A given measurement convention converts the sequence into a rate series. Membership requires records for all 3 components. Operational recastings are fixed across all reported evaluations. Systems without an identifiable substrate or with only aggregate observations lacking substrate attribution are excluded from analyses (Supplementary Table~\ref{tab:si-corpus}).

\subsection*{Data and Corpus Composition}
The corpus contained 25 systems across 5 broad domains: Earth \& Physical Sciences, Life Sciences \& Medicine, Technology \& Information, Trade \& Institutions, and Transportation \& Infrastructure. Each of the 5 broad domains contained 5 individual disciplines (Supplementary Table~\ref{tab:si-corpus}; \hyperref[sec:si-note-processing]{Supplementary Note 2}). Network substrates ranged from 5 to 10,356,390 nodes and from 10 to 7,220,486,175 edges. Maximum rates ranged from 3 to 1,323,743 observations per day. Observation windows ranged from 30 to 8,014 days. Measurements shared no units prior to encoding. Each system contributed exactly one record containing its graph invariants $x$, process signatures $z$, and maximum rate $r_{\max}$.

\subsection*{Network Substrate Construction}
Network substrates were represented as static, unweighted, and undirected graphs. They were strictly held fixed across their observation windows. Nodes represented units. Edges represented native connections between nodes or constructed associations. Directed and weighted edges were converted to undirected and unweighted edges. Multi-edges and self-loops were removed. Isolated nodes were retained. Each network substrate was formally defined as a simple graph $G$:
\begin{equation}
    G \;=\; (V, E),
    \qquad
    E \subseteq \binom{V}{2}
\label{eq:graph}
\end{equation}
Let $\mathcal{O}$ denote the observations from which the dynamical process $S$ was constructed. An attribution map assigned each observation $o \in \mathcal{O}$ a nonempty set $a(o) \subseteq V \cup E$, where $V$ and $E$ denote the node and edge sets, respectively. Interaction logs formed unions of observed pairs or complete bipartite graphs over documented entity types. There were 4 network substrates (college, email, mooc, overflow) defined as complete bipartite graphs $K_{m,n}$ by sizes $m$ and $n$ (\hyperref[sec:si-note-bipartite]{Supplementary Note 1}). Each $G$ served as the fixed network substrate to which observations of the system's dynamical process were attributed.

\subsection*{Dynamical Process Construction}
Dynamical processes were constructed from empirical observations.
For continuous records, measurements were discretized using system-specific thresholds (Supplementary Table~\ref{tab:si-corpus}). Event counts were aggregated into uniform daily intervals using UTC timestamps or consecutive 24-hour periods. Note the exception of 1 system (celegans) where developmental transitions were reconstructed from published timings using a 30-day dauer duration. Each dynamical process was formally defined as a chronologically ordered discrete daily rate series $S$:
\begin{equation}
    S \;=\; \left\{ r_{t} \right\}_{t \in T},
    \qquad
    r_{t} \;=\; \frac{n_{t}}{\Delta t},
    \qquad
    \Delta t \;=\; 1\ \mathrm{day}\ (24\ \mathrm{hours})
    \label{eq:series}
\end{equation}

where $n_{t}$ is the event count in interval $t$ and $T$ is the chronologically ordered set of intervals. The number of included intervals was $L = \lvert T \rvert$, with series mean $\bar{r} = L^{-1}\sum_{t \in T} r_t$. Source conventions determined interval inclusion. For 19 systems, $T$ contained only days with at least one recorded event. There were 5 systems that supplied regular snapshots aggregated into consecutive daily intervals, including zero-event days (metrla, pemsbay, montevideo, windmill, and wiki) and 1 system (chickenpox) that computed a daily-average rate at each weekly report date. Each $S$ served as the realization of the system's dynamical process across its observation window.

\subsection*{Maximum Rate Target}
The target value for each system was the log-transformed maximum rate realized across its observation window. The log-transformation compressed the range of maxima. For each rate series $S$ on interval set $T$ (Eq.~\eqref{eq:series}), the maximum rate $r_{\max}$ and log-transformed maximum rate target $y^*$ are formally defined as:
\begin{equation}
    y^{*} \;=\; \ln\!\left(1 + r_{\max}\right),
    \qquad
    r_{\max} \;=\; \max_{t \in T} r_t
\label{eq:maximum}
\end{equation}
Maxima can grow with observation window length \citep{ref58,ref59,ref60}. Observation windows varied across sources and were held at their recorded durations. All frontiers remained conditional on their recorded observational windows (Supplementary Table~\ref{tab:si-corpus}). Each $y^*$ served as the target bounded by the capacity frontier across its observation window.

\subsection*{Encoding and Standardization}
\label{sec:encoding}
Each network substrate was encoded using 21 graph invariants across 5 categories: simple, cohesion, extremal, statistical, and spectral. The mapping $G \to x$ is formally defined as:
\begin{equation}
    x \;=\; \phi(G) \;=\; (x_1, x_2, \ldots, x_{21}) \;\in\; \mathbb{R}^{21}
\label{eq:invariants}
\end{equation}
For the 4 complete bipartite substrates (college, email, mooc, overflow), invariants were evaluated using exact analytical expressions for $K_{m,n}$ to avoid instantiating billions of edges.

Each dynamical process was encoded using 7 process signatures across 5 categories: dependence, dispersion, volatility, trend, and asymmetry. The mapping $S \to z$ is formally defined as:
\begin{equation}
    z \;=\; \psi(S) \;=\; (z_1, z_2, \ldots, z_7) \;\in\; \mathbb{R}^{7}
\label{eq:signatures}
\end{equation}
6 process signatures were scale-free. The Fano factor retained count scale, equaling the series mean $\bar{r}$ multiplied by the squared coefficient of variation, reflecting process dispersion.

$x$ and $z$ were centered by in-sample means and scaled by in-sample population standard deviations, with 0 standard deviations replaced by 1. The standardized vectors are formally defined as:
\begin{equation}
x' \;=\; \frac{x - \mu_x}{\sigma_x},
\qquad
z' \;=\; \frac{z - \mu_z}{\sigma_z}
\label{eq:standardize}
\end{equation}
Means and scales were fit strictly in-sample (Supplementary Tables~\ref{tab:si-features}--\ref{tab:si-signatures}). 

All features met the admission criteria: classical, deterministic, global to the network or series, free of arbitrary thresholds, without external parameters, and finite.

\subsection*{Frontier Learning}
\phantomsection
\label{sec:frontier-learning}
The capacity frontier was learned through a two-stage log-additive formulation. The structural stage (stage 1) was $f_C(x')$, and the process stage (stage 2) was $f_R(z')$. The same estimator type was trained separately in each stage. Hyperparameters were held constant. The estimated frontier value $\widehat{y}^*$ is formally defined as:
\begin{equation}
\widehat{y}^* = F(x', z') = f_C(x'; \tau_C) + f_R(z'; \tau_R)
\label{eq:frontier}
\end{equation}

where log-transformed maximum rates are bounded by this envelope ($y^* \leq \widehat{y}^*$), subject to slack and potential violations.
\clearpage

Stage 1 trained the structural component $f_C(x'; \tau_C)$ to target $y^{*}$ via asymmetric regression at $\tau_C = 0.99$. It placed the learned function near the upper-envelope given the graph invariants \citep{ref29}. Stage 2 trained the process component $f_R(z'; \tau_R)$ to the signed slack $y^{*} - f_C(x'; \tau_C)$ at $\tau_R = 0.50$. Subtracting the process stage's training mean $\mu_R$ fixed its additive offset \citep{ref67}, resulting in the centered contribution $f_R(z'; \tau_R)$ while leaving $f_C(x'; \tau_C)$ unchanged. Summing both stages yielded the frontier estimate $\widehat{y}^*$. $\tau$ denotes the active stage's asymmetry level $\tau_C$ (stage 1) and $\tau_R$ (stage 2).

The linear parametric paradigm included linear quantile regression \citep{ref22} minimizing mean pinball loss with $\ell_1$-penalized slopes ($\alpha = 0.1$), unpenalized intercepts in both stages, and the \texttt{HiGHS 1.8.0} solver. Linear Asymmetric Weighted Least Squares (LAWS) minimized $(2n)^{-1}\sum_{i=1}^{n} w_i e_i^2 + \alpha\lVert\beta\rVert_1$ ($\alpha = 0.1$), with $w_i = \tau$ for $e_i \ge 0$ and $1-\tau$ otherwise \citep{ref36}. Here $e_i$ is the active stage's observed training response minus its prediction, $n$ the training sample size, and $\beta$ the slope vector. Cyclic coordinate descent used zero initial slopes, tolerance $10^{-4}$, and at most 1,000 passes \citep{ref76}. LAWS used an unpenalized stage 1 intercept initialized at the target mean and no stage 2 intercept. The concave-envelope regressor fitted a stage-specific linear quantile score ($\alpha = 0.1$) and interpolated the concave upper-envelope of target values with constant extrapolation beyond endpoints.

The non-linear ensemble paradigm included a quantile random forest with 100 fully grown bootstrap trees, squared-error splits, a minimum leaf size of 1, and all predictors eligible at each split \citep{ref37}. Forest predictions were quantiles of the fitted conditional target distribution. Gradient-boosted quantile regression minimized pinball loss using 100 trees with maximum depth 3, learning rate 0.1, and the full training sample \citep{ref38}. XGBoost minimized pinball loss with histogram tree construction, 100 trees, maximum depth 6, learning rate 0.3, an $\ell_2$ penalty $\lambda = 1$, and full row and feature sampling \citep{ref39}. Each estimator used the quantile assigned to the active stage.

The neural network paradigm included an MLP with pinball loss, an MLP with expectile loss, and an input-convex network with pinball loss \citep{ref40}. Non-input weights in the convex network were made non-negative through absolute values. All networks used hidden layers of 8 and 4 units, ReLU activations \citep{ref80}, dropout of 0.1 \citep{ref77}, and output biases initialized at the active stage's training quantile. Full-batch AdamW optimization used learning rate 0.1, weight decay 0.01, gradient-norm clipping at 5.0, and 1,000 epochs \citep{ref41}. Each stage's predictions were clipped to its own training target range before process centering and summation.

\subsection*{Performance Metrics}
\phantomsection
\label{sec:performance-metrics}
Frontier efficiency was measured using the Efficiency Index (EI). EI is the geometric mean of the non-violation rate $(1 - \mathrm{VR})$, the bounded mean-violation score $\smash{{(1 + \mathrm{MV}/\overline{\lvert y^{*} \rvert})}^{-1}}$, and the bounded mean-slack score $\smash{{(1 + \mathrm{MS}/\overline{\lvert y^{*} \rvert})}^{-1}}$, formally defined as:
\begin{equation}
    \mathrm{EI} \;=\;
    \Big[
        (1 - \mathrm{VR}) \cdot
        \Big{(1 + \mathrm{MV}/\overline{\lvert y^{*} \rvert}\Big)}^{-1} \cdot
        \Big{(1 + \mathrm{MS}/\overline{\lvert y^{*} \rvert}\Big)}^{-1}
    \Big]^{1/3} \in [0, 1]
    \label{eq:ei}
\end{equation}
For $n$ evaluated systems with observed targets $y_i^{*}$ and predicted values $\widehat{y}_i^{*}$, let $\mathcal{V} = \{i : y_i^{*} > \widehat{y}_i^{*}\}$ denote the violators. The Violation Rate (VR) is the fraction of evaluated systems exceeding their predictions, $\mathrm{VR} = \lvert \mathcal{V} \rvert/n$. The Mean Violation (MV) is the mean exceedance among violators, $\mathrm{MV} = \big(\sum_{i \in \mathcal{V}} (y_i^{*} - \widehat{y}_i^{*})\big)/\lvert \mathcal{V} \rvert$ (0 when $\lvert \mathcal{V} \rvert = 0$). The Mean Slack (MS) is the mean non-negative gap between $\widehat{y}_i^{*}$ and $y_i^{*}$ across all evaluated systems, $\mathrm{MS} = n^{-1}\sum_{i=1}^{n} \max(\widehat{y}_i^{*} - y_i^{*}, 0)$. MV and MS are normalized by the sample mean absolute target $\overline{\lvert y^{*} \rvert} = n^{-1}\sum_{i=1}^{n} \lvert y_i^{*} \rvert$. For numerical stability, $10^{-10}$ was added to the normalization denominator and used as a floor for $1 - \mathrm{VR}$. MV and MS are expressed in natural-log units. Note that EI = 1 under perfect frontier efficiency, which requires no violations and no slack (VR = MV = MS = 0).
\clearpage

Frontier consensus was measured using the Consensus Index (CI). CI is the geometric mean of rescaled Spearman rank correlation $\rho_s^{+} = (1 + \rho_s)/2$ \citep{ref42}, top-weighted Rank-Biased Overlap (RBO) \citep{ref43}, and distance correlation ($\mathrm{dCor}$) \citep{ref44}, formally defined as:
\begin{equation}
\mathrm{CI} \;=\;
\Big[
\rho_s^{+} \cdot \mathrm{RBO} \cdot \mathrm{dCor}
\Big]^{1/3} \in [0, 1]
\label{eq:ci}
\end{equation}
$\rho_s$ used average ranks for ties and was set to 0 when either vector was constant, yielding $\rho_s^{+} = 0.5$. RBO evaluated descending rankings with persistence $p_{\mathrm{RBO}} = 0.9$. Ties were resolved by ascending input index. RBO used tail extrapolation from agreement at the last evaluated depth. $\mathrm{dCor}$ used the biased V-statistic sample estimator with Euclidean-distance exponent 1 and 0 for a constant vector. All CI components were clipped $[0, 1]$ against floating point imprecision. CI measured estimation-observation consensus between predicted values $\widehat{y}^*$ and observed targets $y^*$ as well as pairwise model consensus between predicted values $\widehat{y}^*_i$ and $\widehat{y}^*_j$ from distinct estimators $i \neq j$. Note that CI = 1 under perfect frontier consensus, which requires that all three components be in perfect agreement ($\rho_s^{+} = \mathrm{RBO} = \mathrm{dCor} = 1$).

EI and CI serve as paired metrics designed to measure complementary aspects of frontier performance. EI measures boundary containment and envelope tightness by penalizing violations and excessive slack. CI measures system-specific fidelity by measuring monotonic ordering, top-capacity overlap, and statistical dependence. Critically, neither metric is sufficient in isolation. An arbitrarily high and uninformative constant frontier above all observations can achieve a noteworthy EI despite its slack, yet degrades consensus. Conversely, an uncalibrated ranking can achieve high CI while failing to bound the data. A valid capacity frontier must simultaneously demonstrate containment and fidelity.

\subsection*{Resampling and Leakage Control}
\phantomsection
\label{sec:resampling}
Generalization was evaluated using 4 resampling regimes: repeated $5$-Fold-CV, repeated $10$-Fold-CV, domain-level LOGO-CV, and discipline-level LOGO-CV. Repeated $k$-Fold-CV used random reshuffling over 30 repetitions \citep{ref73}. Domain-level LOGO-CV withheld 1 of the 5 broad domains for each domain. Discipline-level LOGO-CV withheld 1 of the 25 disciplines for each discipline \citep{ref68,ref69,ref70,ref74,ref107}. LOGO-CV folds were deterministic. Repetitions were drawn across training seeds $42 + b$ for $b = 0, \ldots, 29$.

Within each fold, standardizers for $x$ and $z$ (Eq.~\eqref{eq:standardize}), both estimators $f_C(.)$ and $f_R(.)$, and the process centering mean $\mu_R$ (Eq.~\eqref{eq:frontier}) were fit \textit{exclusively} on the training set. Estimators were freshly cloned on every fold. At inference time, models $f_C(x')$ and $f_R(z')$ had no access to the target $y^*$ or raw rate series $S$. Recall that $y^*$ was log-transformed, which did not require parameter fitting. Therefore, information from out-of-sample systems did not enter standardization and model training on the original corpus during cross-validation.

Falsified negative-control corpora were generated from the full corpus, and feature-wise sample standard deviations for additive-noise perturbations were computed across the full corpus before cross-validation. These corpus-wide operations specified the negative controls and scales related to additive-noise. Across all cross-validated stress tests, standardizers and estimators were fit \textit{exclusively} on the corresponding training set.

\subsection*{Perturbation, Falsification, Ablation}
\phantomsection
\label{sec:stress}
Network substrate perturbations at intensity $\epsilon$ modified $G = (V, E)$ (Eq.~\eqref{eq:graph}) across prespecified parameter grids and protocols (Supplementary Table~\ref{tab:si-protocol}). Rewiring used $\widetilde{G} = \mathcal{R}_q(G)$, where $\mathcal{R}_q$ performed $q = \max(1, \lfloor \epsilon\lvert E \rvert \rfloor)$ degree-preserving trials on a simple graph \citep{ref56}. Densification used $\widetilde{G} = (V, E \cup A)$, where $A$ was sampled uniformly without replacement from $\binom{V}{2} \setminus E$, with $\lvert A \rvert = \min\{\lfloor \epsilon\lvert E \rvert \rfloor, \binom{\lvert V \rvert}{2} - \lvert E \rvert\}$ \citep{ref106}. Node removal used the induced subgraph $\widetilde{G} = G[V \setminus U]$, where $U$ was sampled uniformly without replacement from $V$ and $\lvert U \rvert = \lfloor \epsilon\lvert V \rvert \rfloor$ \citep{ref104,ref105}. Graph invariants were recomputed as $\widetilde{x} = \phi(\widetilde{G})$ (Eq.~\eqref{eq:invariants}). For the 4 complete bipartite substrates, node removal used exact closed-form invariants. Rewiring and densification used deterministic approximations (\hyperref[sec:si-note-bipartite]{Supplementary Note 1}).

Dynamical process perturbations modified the rate series through power-law scaling, trailing-mean smoothing, and bootstrapping. For these expressions, $r_{\ell}$ denotes the rate at chronological position $\ell = 1, \ldots, L$, where $L = \lvert T \rvert$. Scaling used $\widetilde{r}_{\ell} = (r_{\ell} + 1)^{\gamma} - 1$. Smoothing used $\widetilde{r}_{\ell} = (\sum_{h=\max(1,\ell-w+1)}^{\ell} r_h)/\min(w,\ell)$, shortening the $w$-interval window at the beginning of the series. Both methods were deterministic. Bootstrapping drew independent indices $I_h \sim \operatorname{Unif}\{1, \ldots, L\}$ for $h = 1, \ldots, k$ \citep{ref75}, with $k = \max(1, \operatorname{round}(sL))$ and halves rounded to even. The first $k$ intervals were $\widetilde{r}_{\ell} = r_{I_{\ell}}$ for $\ell \le k$. The remaining intervals were $\widetilde{r}_{\ell} = r_{I_{J_{\ell}}}$ for $\ell > k$, with independent $J_{\ell} \sim \operatorname{Unif}\{1, \ldots, k\}$. Process signatures were recomputed as $\widetilde{z} = \psi(\widetilde{S})$ (Eq.~\eqref{eq:signatures}), where $\widetilde{S}$ was the modified series. All perturbations retained the original targets.

Graph invariants and process signatures were also perturbed through additive noise, multiplicative jitter, and coordinate subsetting. These methods were applied separately to $x$ and $z$ prior to standardization. For either vector $v \in \{x, z\}$, additive noise used $\widetilde{v}_j = v_j + \varepsilon_j$, with $\varepsilon_j \sim \mathcal{N}(0, (s\sigma_j)^2)$ and $\sigma_j$ the corpus-wide sample standard deviation of coordinate $j$. Jitter used $\widetilde{v}_j = \max\{v_j(1 + \eta_j), 0\}$ for originally non-negative $v_j$ and $\widetilde{v}_j = v_j(1 + \eta_j)$ otherwise, with $\eta_j \sim \mathcal{N}(0, s^2)$. Masking used $\widetilde{v}_j = v_j\mathbf{1}[j \in M]$, where $M$ was a uniformly sampled subset of retained coordinates with $\lvert M \rvert = \lfloor \rho p \rfloor$. Here $\rho$ was the retention fraction, $p = 21$ for $x$ or $p = 7$ for $z$. Each system used the same seed within a realization, sharing coordinate-wise noise and jitter draws and the masked subset across systems.

The 3 falsification procedures deliberately destroyed the mapping $(x', z') \to y^*$. Let $i = 1, \ldots, 25$ index systems and $D_i$ denote their broad domains. Target permutation used $\widetilde{y}_i^{*} = y_{\pi(i)}^{*}$, where $\pi$ was a permutation satisfying $D_{\pi(i)} \ne D_i$, preserving each system's original $x$ and $z$ and the target's marginal distribution. For either raw vector $v \in \{x, z\}$, random generation independently sampled $\widetilde{v}_{ij} \sim \mathcal{U}(a_j, b_j)$, where $a_j = \min_h v_{hj}$ and $b_j = \max_h v_{hj}$ across the corpus. Vector generation resampled empirical coordinates with replacement \citep{ref75} and added Gaussian noise as $\widetilde{v}_{ij} = v_{B_{ij},j} + \xi_{ij}$, with independent $B_{ij} \sim \operatorname{Unif}\{1, \ldots, 25\}$ and $\xi_{ij} \sim \mathcal{N}(0, \nu_j^2)$. Here $\nu_j$ was the coordinate's corpus-wide population standard deviation, floored at $10^{-10}$. Both feature-generation procedures retained each system's original target.

The 4 ablations compared alternative specifications with the log-additive decomposition using the original training folds and standard estimator protocols. Invariants only used $F_{\mathrm{Inv}}(x') = f_{\mathrm{Inv}}(x'; \tau_C)$. Signatures only used $F_{\mathrm{Sig}}(z') = f_{\mathrm{Sig}}(z'; \tau_C)$. Joint features used $F_{\mathrm{Joint}}(x', z') = f_{\mathrm{Joint}}([x', z']; \tau_C)$, where brackets denote vector concatenation. These ablations used single-stage fits at $\tau_C = 0.99$. Interaction terms formed a 147-coordinate vector $h = {(x_j z_k)}_{j=1,\ldots,21\,;\,k=1,\ldots,7}$ from raw coordinates and standardized it within each training fold to $h'$. The alternative frontier was $F_{\mathrm{Int}}(x', z', h') = f_{C,\mathrm{Int}}(x'; \tau_C) + f_{R,\mathrm{Int}}([z', h']; \tau_R)$, with both stages refit and $f_{R,\mathrm{Int}}$ centered (Eq.~\eqref{eq:frontier}).

\subsection*{Evaluation and Statistical Inference}
\phantomsection
\label{sec:si-stat-conventions}
Inference used 45 estimator-by-domain comparisons summarized over 30 repetitions. Within each domain, EI was computed using seed-averaged predictions for resampling equivalence and falsification. Perturbation and ablation used repetition means in both the original and tested conditions. CI used seed-averaged predictions except for stochastic perturbations, where metrics were averaged after computing each paired training seed and realization.

Paired differences $\Delta\mathrm{EI}$ and $\Delta\mathrm{CI}$ were computed as test comparison minus baseline and summarized by their median. Baselines were domain-level LOGO-CV for resampling equivalence, the original condition for perturbation and falsification, and the log-additive decomposition for ablation. All comparisons used One-Sided Wilcoxon Signed-Rank tests \citep{ref45}, assuming symmetric paired-difference distributions with location parameter $\theta$. Inference was conditional on the given panel whose shared estimators and cross-validation folds induced dependence that may affect $p$-value calibration.

For a margin $\delta > 0$, resampling equivalence and perturbation stability used TOST~\citep{ref47} of $H_{01}: \theta \le -\delta$ and $H_{02}: \theta \ge \delta$. Equivalence required rejection of both nulls with the TOST $p$-value taken as the larger of the 2 one-sided values. Falsification tested degradation ($H_0: \theta \ge 0$ versus $H_1: \theta < 0$). Ablation tested non-inferiority of the log-additive decomposition to each alternative \citep{ref78} ($H_0: \theta \ge \delta$ versus $H_1: \theta < \delta$).

All margins $\delta$ were empirically derived based on interquartile ranges (IQR) rounded down 2 decimal places (Supplementary Tables~\ref{tab:si-transfer-equivalence}--\ref{tab:si-perturb} and \ref{tab:si-ablation}). Holm-Bonferroni adjustments (Holm-adjusted $p$, denoted $p_{\mathrm H}$) \citep{ref46} used a significance threshold of $\alpha = 0.05$ separately for EI and CI within 5 families: resampling, perturbation, falsification, simple ablations, and complex ablations.

Rank-biserial effect sizes were computed as $r = (W_{+} - W_{-})/(W_{+} + W_{-})$ from nonzero, unshifted paired differences \citep{ref71}. Here $W_{+}$ and $W_{-}$ are the sums of ranks of absolute differences with positive and negative signs, respectively, with average ranks for ties.

Full-corpus consensus tests were descriptive (Fig.~\ref{fig:consensus}). Each estimator's predictions were averaged across training seeds. For stochastic perturbations, realizations were paired before computing estimation-observation and pairwise model CI. Pairwise model consensus was the median CI across the 36 distinct estimator pairs. Original data fits were frozen for evaluation on the highest intensity perturbed corpora and the falsified corpora. Ablations were refit on the full original corpus.

\subsection*{Implementation and Reproducibility}
\label{sub-sec:imp-rep}

All analyses were implemented in \texttt{Python 3.11.9} \citep{ref86,ref102}. All dependency versions were pinned in the repository's \texttt{requirements.txt}. Graphs, processes, and encodings used \texttt{python-igraph 1.0.0} \citep{ref48,ref93}, \texttt{NumPy 2.3.5} \citep{ref81,ref96}, \texttt{SciPy 1.16.3} \citep{ref82,ref97}, and \texttt{pandas 3.0.0} \citep{ref83,ref84}. Urban street network construction used \texttt{networkx 3.5} \citep{ref91,ref98} and \texttt{OSMnx 2.0.6} \citep{ref118,ref119}.

Public dataset retrieval was conducted using \texttt{requests 2.32.5} \citep{ref120}. Temporal graph datasets were loaded using \texttt{torch-geometric 2.7.0} \citep{ref89,ref99} and \texttt{torch\_geometric\_temporal 0.56.2} \citep{ref90,ref100}. Observational records for gravitational wave and meteorological systems were retrieved using \texttt{gwosc 0.8.1} and \texttt{meteostat 1.7.6}, respectively \citep{ref116,ref117}.

Estimators used \texttt{scikit-learn 1.8.0} \citep{ref72,ref94}, \texttt{sklearn-quantile 0.1.1} \citep{ref87}, \texttt{XGBoost 3.1.1} \citep{ref39,ref95}, and \texttt{PyTorch 2.9.0} \citep{ref85,ref115}. Distance correlation was computed using \texttt{dcor 0.7} \citep{ref88}, and scientific figures were generated with \texttt{matplotlib 3.9.2} \citep{ref92,ref101}. Unspecified settings followed library defaults.

All stochastic procedures started from random seed 42 with repeat indices applied as fixed numerical offsets. Random permutations, cross-validation folds, and perturbation realizations originated from seeded generators. 

\texttt{PyTorch 2.9.0} deterministic algorithms were executed in warn-only mode and execution defaulted to CUDA when GPU resources were available and CPU otherwise. Results reproduced with the code repository, corpus, and environment specifications are subject to floating-point imprecision.
\clearpage

\section*{Data Availability}
\label{sec:data}
The corpus was collated from publicly available data sources (Supplementary Table~\ref{tab:si-corpus}). The processed records used in all analyses are distributed with the source code repository at: \\
\url{https://github.com/nickkunz/network-capacity}.

\section*{Code Availability}
\label{sec:code}
The scripts for data processing, frontier estimation, evaluation, perturbation, falsification, and ablation, along with the notebooks that reproduce every figure and table are also distributed with the source code repository at: \url{https://github.com/nickkunz/network-capacity}.

\section*{Acknowledgements}
\label{sec:acknowledge}
We would like to thank F. Parise, W. Xu, N. de Santi, and S. Visvesvaran for their early feedback, discussions, and encouragement.

\section*{Author Contributions}
\label{sec:authors}
N. Kunz initiated and developed the concept, methods, implementation, testing, documentation, and manuscript. H. O. Gao initiated and supervised the development of the investigation and manuscript. The authors have agreed to the submitted version of this manuscript.

\section*{Funding}
\label{sec:funding}
This work was supported by Cornell University and Gao Labs.

\section*{Competing Interests}
\label{sec:competing}
The authors declare no competing interests.

\section*{Additional Information}
\label{sec:additional}
\textbf{Supplementary information} is in Supplementary Tables~\ref{tab:si-corpus}--\ref{tab:si-protocol} and Supplementary Notes~\hyperref[sec:si-note-bipartite]{1}--\hyperref[sec:si-note-processing]{2}.

\textbf{Correspondence and requests for materials} should be addressed to N. Kunz.

\newpage
\bibliographystyle{unsrtnat}
\bibliography{ref}

\clearpage
\begingroup
\renewcommand{\tablename}{Supplementary Table}
\renewcommand{\theHtable}{supp.\arabic{table}}
\setcounter{table}{0}
\captionsetup[table]{font=footnotesize,labelfont=bf,labelsep=period,justification=raggedright,singlelinecheck=false}
\setlength{\tabcolsep}{3pt}
\renewcommand{\arraystretch}{1.15}
\footnotesize

\section*{Supplementary Tables}
\begingroup
\captionsetup{hypcap=false}
\captionof{table}{Corpus composition, definitions, sources, and date provenance.}
\label{tab:si-corpus}
\endgroup

\begin{center}
\rotatebox{90}{%
\begin{minipage}{.92\textheight}
\scriptsize
\setlength{\tabcolsep}{2pt}
\begin{tabular}{@{}
    >{\raggedright\arraybackslash}p{\dimexpr.08\linewidth-2\tabcolsep\relax}
    >{\raggedright\arraybackslash}p{\dimexpr.10\linewidth-2\tabcolsep\relax}
    >{\raggedright\arraybackslash}p{\dimexpr.10\linewidth-2\tabcolsep\relax}
    >{\raggedright\arraybackslash}p{\dimexpr.10\linewidth-2\tabcolsep\relax}
    >{\raggedright\arraybackslash}p{\dimexpr.10\linewidth-2\tabcolsep\relax}
    >{\raggedright\arraybackslash}p{\dimexpr.10\linewidth-2\tabcolsep\relax}
    >{\raggedright\arraybackslash}p{\dimexpr.10\linewidth-2\tabcolsep\relax}
    >{\raggedright\arraybackslash}p{\dimexpr.08\linewidth-2\tabcolsep\relax}
    >{\raggedright\arraybackslash}p{\dimexpr.105\linewidth-2\tabcolsep\relax}
    >{\raggedright\arraybackslash}p{\dimexpr.135\linewidth-2\tabcolsep\relax}@{}}
\toprule
\textbf{System} & \textbf{Domain} & \textbf{Discipline} & \textbf{Nodes} & \textbf{Edges} & \textbf{Event} & \textbf{Target} & \textbf{Time (Days)} & \textbf{Source} & \textbf{Notes} \\
\midrule
\textbf{\textit{gwosc}} & Earth \& Physical Sciences & Cosmology &
Gravitational-\allowbreak wave detectors (H1, L1, V1, K1, G1). & Complete graph among detectors. &
GWTC event records from any GWTC catalog attributed to at least one corpus detector, including the marginal and auxiliary catalogs. & Maximum daily records $r_{\max} = 3$, $y^* = 1.386$. & 2015-09-14 to 2024-01-09 ($n = 3{,}040$). &
GWOSC event API (\ref{src:S1}, \ref{src:S24}, \ref{src:S25}, \ref{src:S26}, \ref{src:S27}, \ref{src:S30}). & Catalog articles published in 2019, 2021, 2023, and 2024. GWTC-4.0 first released 26 August 2025 and its results article was published in 2026. Fixed event cutoff 2024-01-09. \\
\addlinespace[3.5pt]
\textbf{\textit{river}} & Earth \& Physical Sciences & Hydrology &
USGS stream gauges on the Colorado River (HUC 15010001, 15010002, 15010005). & Non-crossing path among gauges. &
Peaks of discharge (parameter 00060) above the site-wise 99th percentile. & Maximum daily discharge-peak exceedances $r_{\max} = 9$, $y^* = 2.303$. & 2014-11-12 to 2023-08-01 ($n = 3{,}185$). &
USGS NWIS site and instantaneous-\allowbreak values services (\ref{src:S2}). & NWIS dataset record published 1994. API query cutoff 2024-12-31. \\
\addlinespace[3.5pt]
\textbf{\textit{auger}} & Earth \& Physical Sciences & Physics &
Pierre Auger Observatory surface-\allowbreak detector stations. & Reciprocal 6-nearest neighbours within $1.1\times$ array spacing (1,500 m and 750 m arrays). &
Surface-\allowbreak detector air-shower events. & Maximum daily air-shower events $r_{\max} = 66$, $y^* = 4.205$. & 2004-01-03 to 2018-09-01 ($n = 5{,}356$). &
Auger Open Data station map and Zenodo record 10488964 (\ref{src:S3}). & Open Data version 3 released 2024. Observations span 2004--2018. \\
\addlinespace[3.5pt]
\textbf{\textit{seismic}} & Earth \& Physical Sciences & Geology &
IRIS IU network stations. & Complete graph among stations. &
Reviewed USGS events, 2023. & Maximum daily reviewed events $r_{\max} = 1{,}129$, $y^* = 7.030$. & 2023-01-01 to 2023-12-30 ($n = 364$). &
IRIS FDSN station service (\ref{src:S4}) and USGS FDSN event service (\ref{src:S5}). & GSN dataset record published 1988. ComCat record published 2017. Event-query cutoff 2023-12-31. \\
\addlinespace[3.5pt]
\textbf{\textit{rain}} & Earth \& Physical Sciences & Meteorology &
Meteostat hourly stations in Laos. & Delaunay triangulation of station coordinates. &
Station-hours with precipitation $>1$ mm. & Maximum daily wet station-hours $r_{\max} = 46$, $y^* = 3.850$. & 2024-01-12 to 2024-03-30 ($n = 79$). &
Meteostat (\ref{src:S6}). & Mutable service accessed 1 September 2026. Query cutoff 2024-03-31. \\
\addlinespace[3.5pt]
\textbf{\textit{celegans}} & Life Sciences \& Medicine & Biology &
\textit{C. elegans} developmental stages. & Known stage-to-stage transitions (normal, dauer, and recovery pathways), entered undirected. &
Developmental transitions at published mean times. & Maximum daily developmental transitions $r_{\max} = 8$, $y^* = 2.197$. & Days 0--33 ($n = 34$). &
Cassada \& Russell, \textit{Dev. Biol.} \textbf{46}, 326--342 (1975) (\ref{src:S7}), and Golden \& Riddle, \textit{Dev. Biol.} \textbf{102}, 368--378 (1984) (\ref{src:S8}). & Timing studies published 1975 and 1984. Days are developmental-age indices, not calendar dates. \\
\midrule
\multicolumn{10}{r@{}}{\textit{Continued on next page}}\\
\end{tabular}
\end{minipage}}
\end{center}

\begin{landscape}
\begingroup
\scriptsize
\setlength{\tabcolsep}{2pt}
\addtocounter{table}{-1}
\renewcommand{\theHtable}{supp.1.cont}
\begin{longtable}{@{}
    >{\raggedright\arraybackslash}p{\dimexpr.08\linewidth-2\tabcolsep\relax}
    >{\raggedright\arraybackslash}p{\dimexpr.10\linewidth-2\tabcolsep\relax}
    >{\raggedright\arraybackslash}p{\dimexpr.10\linewidth-2\tabcolsep\relax}
    >{\raggedright\arraybackslash}p{\dimexpr.10\linewidth-2\tabcolsep\relax}
    >{\raggedright\arraybackslash}p{\dimexpr.10\linewidth-2\tabcolsep\relax}
    >{\raggedright\arraybackslash}p{\dimexpr.10\linewidth-2\tabcolsep\relax}
    >{\raggedright\arraybackslash}p{\dimexpr.10\linewidth-2\tabcolsep\relax}
    >{\raggedright\arraybackslash}p{\dimexpr.08\linewidth-2\tabcolsep\relax}
    >{\raggedright\arraybackslash}p{\dimexpr.105\linewidth-2\tabcolsep\relax}
    >{\raggedright\arraybackslash}p{\dimexpr.135\linewidth-2\tabcolsep\relax}@{}}
\multicolumn{10}{@{}l@{}}{\textbf{Supplementary Table~\ref{tab:si-corpus}.} Corpus composition and provenance \textit{(continued).}}\\[4pt]
\toprule
\textbf{System} & \textbf{Domain} & \textbf{Discipline} & \textbf{Nodes} & \textbf{Edges} & \textbf{Event} & \textbf{Target} & \textbf{Time (Days)} & \textbf{Source} & \textbf{Notes} \\
\midrule
\endfirsthead
\multicolumn{10}{@{}l@{}}{\textbf{Supplementary Table~\ref{tab:si-corpus}.} Corpus composition and provenance \textit{(continued).}}\\[4pt]
\toprule
\textbf{System} & \textbf{Domain} & \textbf{Discipline} & \textbf{Nodes} & \textbf{Edges} & \textbf{Event} & \textbf{Target} & \textbf{Time (Days)} & \textbf{Source} & \textbf{Notes} \\
\midrule
\endhead
\midrule
\multicolumn{10}{r@{}}{\textit{Continued on next page}}\\
\endfoot
\bottomrule
\endlastfoot
\textbf{\textit{crop}} & Life Sciences \& Medicine & Ecology &
Crops and pollinator taxa. & Observed crop--pollinator interactions. &
Daily flower-visitor census visits from primary study records. & Maximum daily census visits $r_{\max} = 1{,}266$, $y^* = 7.144$. & 2004-05-07 to 2012-05-29 ($n = 2{,}945$). &
CropPol repository (Rader et al. primary census records) (\ref{src:S9}, \ref{src:S31}). & Database article published 2022. Observations span 2004--2012. \\
\addlinespace[5pt]
\textbf{\textit{chickenpox}} & Life Sciences \& Medicine & Epidemiology &
Hungarian counties. & County adjacency from the Chickenpox dataset. &
Weekly county case counts summed across counties, divided by 7 to a daily average. & Maximum daily-average cases $r_{\max} = 378$, $y^* = 5.938$. & 2005-01-03 to 2014-12-29 ($n = 3{,}648$). &
Rozemberczki et al. Chickenpox dataset and UCI Hungarian chickenpox cases (\ref{src:S10}). & UCI dataset released 2021. Observations span 2005--2014. Source reports weekly totals, converted to daily averages. \\
\addlinespace[5pt]
\textbf{\textit{faers}} & Life Sciences \& Medicine & Pharmacology &
Medicinal products and MedDRA reaction terms co-reported with imatinib. & Drug--reaction co-occurrence. &
Unique FAERS case receipts. & Maximum daily case receipts $r_{\max} = 45$, $y^* = 3.829$. & 2004-01-23 to 2025-12-31 ($n = 8{,}014$). &
openFDA drug-event API, query IMATINIB (\ref{src:S11}). & Mutable API retrieved 13 March 2026. Case-receipt cutoff 2025-12-31. \\
\addlinespace[5pt]
\textbf{\textit{epilepsy}} & Life Sciences \& Medicine & Neurology &
19 scalp electrodes (10--20 system). & Anatomical neighbours on the 10--20 montage. &
Annotated seizures in CHB-MIT (chb01--chb24). & Maximum daily annotated seizures $r_{\max} = 27$, $y^* = 3.332$. & Relative recording days 0--76 ($n = 77$). &
PhysioNet CHB-MIT (\ref{src:S12}). & PhysioNet version 1.0.0 released 2010. Calendar dates are unavailable, so recording order is expressed as relative days. \\
\addlinespace[5pt]
\textbf{\textit{federal}} & Trade \& Institutions & Government &
Awarding agencies and recipients. & Agency--\allowbreak recipient contract pairs. &
Waterfowl-\allowbreak related contract start dates. & Maximum daily contract starts $r_{\max} = 17$, $y^* = 2.890$. & 2011-01-19 to 2024-12-30 ($n = 5{,}095$). &
USAspending API, keyword waterfowl (\ref{src:S13}). & Mutable API retrieved 25 August 2026. Award-start cutoff 2024-12-31. \\
\addlinespace[5pt]
\textbf{\textit{mooc}} & Trade \& Institutions & Education &
Students and course actions. & Complete bipartite graph between students and actions. &
Student--action timestamps. & Maximum daily student actions $r_{\max} = 18{,}213$, $y^* = 9.810$. & Days 0--29 ($n = 30$). &
SNAP act-mooc (\ref{src:S14}). & Dataset article published 2019. Timestamps reindexed to relative days. \\
\addlinespace[5pt]
\textbf{\textit{amazon}} & Trade \& Institutions & Commerce &
Reviewers and book products. & Reviewer--\allowbreak product review pairs. &
Review timestamps. & Maximum daily reviews $r_{\max} = 37{,}293$, $y^* = 10.527$. & 1996-05-20 to 2014-07-23 ($n = 6{,}639$). &
SNAP Amazon product graph, Books ratings (\ref{src:S15}). & Dataset article published 2015. Reviews span 1996--2014. \\
\addlinespace[5pt]
\textbf{\textit{bitcoin}} & Trade \& Institutions & Cryptocurrency &
OTC users and integer rating values. & User--\allowbreak rating--\allowbreak user incidences over the complete Bitcoin OTC history. &
Rating timestamps. & Maximum daily ratings $r_{\max} = 319$, $y^* = 5.768$. & 2010-11-08 to 2016-01-25 ($n = 1{,}905$). &
PyTorch Geometric BitcoinOTC (\ref{src:S16}). & Dataset article published 2016. Ratings span 2010--2016. \\
\addlinespace[5pt]
\textbf{\textit{world}} & Trade \& Institutions & International Development &
World Bank lending instruments and countries. & Instrument--\allowbreak country eligibility. &
Project board-approval dates. & Maximum daily project approvals $r_{\max} = 12$, $y^* = 2.565$. & 2014-01-14 to 2024-12-20 ($n = 3{,}994$). &
World Bank projects and country APIs (\ref{src:S17}). & Mutable APIs retrieved 29 September 2026. Approval-year cutoff 2024. \\
\addlinespace[5pt]
\textbf{\textit{wiki}} & Technology \& Information & Knowledge Graphs &
Wikipedia mathematics pages. & Page--page relations from WikiMaths snapshots. &
Daily summed page activity. & Maximum daily summed page activity $r_{\max} = 1{,}323{,}743$, $y^* = 14.096$. & 2019-03-16 to 2021-03-15 ($n = 731$). &
UCI Wikipedia Math Essentials and PyTorch Geometric Temporal WikiMaths (\ref{src:S18}). & UCI dataset released 2021. Activity spans 2019--2021. \\
\addlinespace[5pt]
\textbf{\textit{jodie}} & Technology \& Information & Interaction Networks &
Wikipedia users and pages. & User--page interactions. &
Interaction timestamps. & Maximum daily interactions $r_{\max} = 6{,}087$, $y^* = 8.714$. & Days 0--30 ($n = 31$). &
PyTorch Geometric JODIE, Wikipedia (\ref{src:S14}). & Dataset article published 2019. Timestamps reindexed to relative days. \\
\addlinespace[5pt]
\textbf{\textit{overflow}} & Technology \& Information & Online Forums &
Stack Overflow users and calendar days. & Complete bipartite graph between users and days. &
Comment, answer, and other timed interactions. & Maximum daily timed interactions $r_{\max} = 52{,}446$, $y^* = 10.868$. & 2008-08-01 to 2016-03-06 ($n = 2{,}775$). &
SNAP sx-stackoverflow (\ref{src:S19}). & Dataset article published 2017. Interactions span 2008--2016. \\
\addlinespace[5pt]
\textbf{\textit{email}} & Technology \& Information & Email Exchanges &
European research-\allowbreak institution email accounts and days. & Complete bipartite graph between accounts and days. &
Email timestamps. & Maximum daily emails $r_{\max} = 2{,}270$, $y^* = 7.728$. & Days 0--803 ($n = 804$). &
SNAP email-\allowbreak Eu-\allowbreak core-\allowbreak temporal (\ref{src:S19}). & Dataset article published 2017. Timestamps reindexed to relative days. \\
\addlinespace[5pt]
\textbf{\textit{college}} & Technology \& Information & Social Networks &
UC Irvine college-message users and days. & Complete bipartite graph between users and days. &
Private-message timestamps. & Maximum daily private messages $r_{\max} = 2{,}678$, $y^* = 7.893$. & 2004-04-15 to 2004-10-26 ($n = 195$). &
SNAP CollegeMsg (\ref{src:S20}). & Dataset article published 2009. Messages were recorded in 2004. \\
\addlinespace[5pt]
\textbf{\textit{idling}} & Transportation \& Infrastructure & Mobility &
Halifax drive-network nodes from the unsimplified OSM graph. & Street segments (OSM drive graph, undirected). &
Vehicle-idling records. & Maximum daily idling records $r_{\max} = 732$, $y^* = 6.597$. & 2025-06-02 to 2025-09-02 ($n = 93$). &
OpenStreetMap via OSMnx, query Halifax, Canada (\ref{src:S21}, \ref{src:S22}), and local idling CSVs (\ref{src:S29}). & Data preprint first posted 2024 and revised 2025. Observations are from 2025. OSMnx article published 2025. \\
\addlinespace[5pt]
\textbf{\textit{windmill}} & Transportation \& Infrastructure & Energy &
Wind turbines. & Complete graph among turbines. &
Raw-production on/off transitions (threshold $10^{-6}$). & Maximum daily turbine transitions $r_{\max} = 7$, $y^* = 2.079$. & Days 0--727 ($n = 728$). &
PyTorch Geometric Temporal Windmill\allowbreak Output\allowbreak Large \citep{ref90}. & Loader article published 2021. Source supplies snapshot order rather than calendar dates. \\
\addlinespace[5pt]
\textbf{\textit{metrla}} & Transportation \& Infrastructure & Traffic &
METR-LA loop detectors. & Detector adjacency from METR-LA snapshots. &
Onset of congestion (raw speed at or below the detector-wise 1st percentile). & Maximum daily congestion onsets $r_{\max} = 675$, $y^* = 6.516$. & 2012-03-01 to 2012-06-27 ($n = 119$). &
PyTorch Geometric Temporal METRLA \citep{ref90} (\ref{src:S23}). & Benchmark article published 2018 and loader article 2021. Sensor observations are from 2012. \\
\addlinespace[5pt]
\textbf{\textit{pemsbay}} & Transportation \& Infrastructure & Traffic II &
PEMS-BAY loop detectors. & Detector adjacency from PEMS-BAY snapshots. &
Onset of congestion (raw speed at or below the detector-wise 1st percentile). & Maximum daily congestion onsets $r_{\max} = 634$, $y^* = 6.454$. & 2017-01-01 to 2017-06-29 ($n = 180$). &
PyTorch Geometric Temporal PEMS-BAY \citep{ref90} (\ref{src:S23}). & Benchmark article published 2018 and loader article 2021. Sensor observations are from 2017. \\
\addlinespace[5pt]
\textbf{\textit{montevideo}} & Transportation \& Infrastructure & Ridership &
Montevideo bus stops. & Stop adjacency from MontevideoBus snapshots. &
Stop-hours with standardized inflow above the global 99th percentile. & Maximum daily exceedance stop-hours $r_{\max} = 256$, $y^* = 5.549$. & Days 0--29 ($n = 30$). &
PyTorch Geometric Temporal MontevideoBus \citep{ref90} (\ref{src:S28}). & Data recorded in October 2020. Loader article published 2021. Dates represented as relative days. \\
\end{longtable}
\par\smallskip
\begin{minipage}{\linewidth}
\footnotesize
Each row represents one N3S, with observations aggregated to $r_t$ (Eq.~\eqref{eq:series}) and targets $r_{\max}$ and $y^* = \ln(1 + r_{\max})$ (Eq.~\eqref{eq:maximum}). Metrla, pemsbay, montevideo, windmill, and wiki provide complete daily series, including zero-event days. Chickenpox provides weekly county totals, summed and divided by 7, with one daily-average value per report date; the other 24 systems use daily event counts. The remaining 19 series include only days with recorded events. Windows follow native recording periods, expressed as dates or relative day indices; $n$ is the inclusive span in days, not necessarily the series length $L$. The 7 API systems (gwosc, river, seismic, rain, faers, federal, and world) also use fixed script query dates, with end cutoffs in the source notes. Durations were not standardized, and maxima remain conditional on their windows. Observation dates differ from source release or publication dates; source notes specify versions and relative-date conventions. Additional processing conventions are provided in the supplement (\hyperref[sec:si-note-processing]{Supplementary Note 2}). Citations S1--S31 refer to the supplementary sources (\nameref{sec:si-sources}).
\end{minipage}
\endgroup
\end{landscape}

\clearpage
\begin{table}[H]
\centering
\footnotesize
\captionsetup{position=top,skip=12pt}
\caption{Domain transfer and baseline frontier performance.}
\label{tab:si-transfer}
\begin{tabularx}{\linewidth}{@{}lXrrrr@{}}
\toprule
Evaluation & Method & Median EI [Q1--Q3] & VR & MV & MS \\
\midrule
Transfer & Domain (LOGO-CV) & 0.75 [0.74--0.77] & 0.04 & 0.08 & 6.86 \\
 & Discipline (LOGO-CV) & 0.73 [0.72--0.76] & 0.00 & 0.00 & 6.12 \\
\addlinespace
Baseline & Random ($5$-Fold-CV) & 0.77 [0.76--0.78] & 0.02 & 0.06 & 6.86 \\
 & Random ($10$-Fold-CV) & 0.76 [0.75--0.78] & 0.01 & 0.02 & 6.87 \\
\bottomrule
\end{tabularx}
\par\vspace{12pt}
\begin{minipage}{\linewidth}
\footnotesize
Entries are medians of estimator-level means over folds or groups and 30 repeats (seeds 42--71), with quartiles across 9 estimators in brackets. Groups, folds, repeats, and estimators are equally weighted. LOGO-CV folds are fixed across repeats; random folds are reshuffled. EI denotes the Efficiency Index (\nameref{sec:performance-metrics},~Eq.~\eqref{eq:ei}), VR the violation rate, MV the mean exceedance among within-fold violators (0 if none), and MS the mean non-negative slack across evaluated systems. MV and MS use natural-log target units. Descriptive summaries only.
\end{minipage}
\end{table}

\begin{table}[H]
\centering
\footnotesize
\captionsetup{position=top,skip=12pt}
\caption{Equivalence of resampling regimes to domain-level LOGO-CV.}
\label{tab:si-transfer-equivalence}
\begin{tabularx}{\linewidth}{@{}Xrrrrrr@{}}
\toprule
\multicolumn{7}{c}{\textbf{Domain Transfer Equivalence}} \\
\cmidrule(lr){1-7}
Evaluation & \shortstack{Median\\$\Delta\mathrm{EI}$} & \shortstack{Rank-biserial\\$r$} & \shortstack{TOST\\$p$} & \shortstack{Holm-adj.\\$p$} & Sig. & Eq. \\
\midrule
Discipline (LOGO-CV) & 0.00 & -0.08 & $<0.001$ & $<0.001$ & *** & Yes \\
Random ($5$-Fold-CV) & 0.00 & 0.40 & $<0.001$ & $<0.001$ & *** & Yes \\
Random ($10$-Fold-CV) & 0.00 & 0.24 & $<0.001$ & $<0.001$ & *** & Yes \\
\midrule
\multicolumn{7}{c}{\textbf{Estimation-Observation Consensus Equivalence}} \\
\cmidrule(lr){1-7}
Evaluation & \shortstack{Median\\$\Delta\mathrm{CI}$} & \shortstack{Rank-biserial\\$r$} & \shortstack{TOST\\$p$} & \shortstack{Holm-adj.\\$p$} & Sig. & Eq. \\
\midrule
Discipline (LOGO-CV) & 0.00 & -0.14 & $<0.001$ & $<0.001$ & *** & Yes \\
Random ($5$-Fold-CV) & -0.01 & -0.32 & $<0.001$ & $<0.001$ & *** & Yes \\
Random ($10$-Fold-CV) & 0.00 & -0.34 & $<0.001$ & $<0.001$ & *** & Yes \\
\bottomrule
\end{tabularx}
\par\vspace{12pt}
\begin{minipage}{\linewidth}
\footnotesize
Each contrast uses 45 paired estimator-by-domain cells (9 estimators $\times$ 5 domains), not 45 independent systems. Out-of-sample predictions were averaged over 30 seeds (42--71) before computing EI and CI within the same 5 broad domains for every regime. EI and CI are defined in the methodology (\nameref{sec:performance-metrics},~Eqs.~\eqref{eq:ei} and~\eqref{eq:ci}). Shifts are medians of paired differences (test comparison $-$ domain-level LOGO-CV), not differences of marginal medians. Rank-biserial $r$ uses nonzero, unshifted differences with average ranks for ties; positive values favor the comparison regime. TOST uses paired one-sided Wilcoxon signed-rank tests on $\Delta+\delta$ and $\Delta-\delta$, assuming a symmetric paired-difference distribution with location $\theta$. The nulls are $\theta\leq-\delta$ and $\theta\geq\delta$; unadjusted $p$ is the larger of the two one-sided values. Margins $\delta_{\mathrm{EI}} = 0.02$ and $\delta_{\mathrm{CI}} = 0.11$ are IQRs across the 9 reference estimator-level means after equal averaging across domains, floored to 2 decimals. Raw differences with magnitude $<10^{-10}$ were zeroed before shifting. Tests used \texttt{SciPy 1.16.3}, \texttt{zero\_method='wilcox'}, \texttt{method='auto'}, and \texttt{correction=False}, discarding zeros after shifting. $p_{\mathrm H}$ is Holm-adjusted across the 3 regime comparisons separately for EI and CI. Eq. requires rejection of both nulls at $p_{\mathrm H}<0.05$. Shared estimators, training systems, and domains make inference conditional on this dependent panel (\nameref{sec:si-stat-conventions}).
\end{minipage}
\end{table}

\clearpage
\begin{table}[H]
\centering
\footnotesize
\captionsetup{position=top,skip=12pt}
\caption{Perturbation effects under the highest test intensity.}
\label{tab:si-perturb}
\setlength{\tabcolsep}{2pt}
\begin{tabularx}{\linewidth}{@{}l>{\raggedright\arraybackslash}Xrrrrrr@{}}
\toprule
\multicolumn{8}{c}{\textbf{Domain Transfer}} \\
\cmidrule(lr){1-8}
Perturbation & Method & \shortstack{Median\\$\Delta\mathrm{EI}$} & \shortstack{Rank-biserial\\$r$} & \shortstack{TOST\\$p$} & \shortstack{Holm-adj.\\$p$} & Sig. & Eq. \\
\midrule
Network & Sample & 0.00 & -0.35 & $<0.001$ & $<0.001$ & *** & Yes \\
 & Rewire & 0.00 & -0.23 & $<0.001$ & $<0.001$ & *** & Yes \\
 & Densify & 0.00 & -0.40 & $<0.001$ & $<0.001$ & *** & Yes \\
\addlinespace
Invariants & Jitter & 0.00 & -0.45 & $<0.001$ & $<0.001$ & *** & Yes \\
 & Noise & 0.00 & -0.60 & $<0.001$ & $<0.001$ & *** & Yes \\
 & Subset & 0.00 & 0.05 & $<0.001$ & $<0.001$ & *** & Yes \\
\addlinespace
Process & Bootstrapping & 0.01 & 0.33 & $<0.001$ & $<0.001$ & *** & Yes \\
 & Scaling & -0.01 & -0.45 & $<0.001$ & $<0.001$ & *** & Yes \\
 & Smoothing & -0.02 & -0.75 & $<0.001$ & $<0.001$ & *** & Yes \\
\addlinespace
Signatures & Jitter & 0.00 & -0.31 & $<0.001$ & $<0.001$ & *** & Yes \\
 & Noise & 0.00 & -0.44 & $<0.001$ & $<0.001$ & *** & Yes \\
 & Subset & -0.01 & -0.61 & $<0.001$ & $<0.001$ & *** & Yes \\
\midrule
\multicolumn{8}{c}{\textbf{Estimation-Observation Consensus}} \\
\cmidrule(lr){1-8}
Perturbation & Method & \shortstack{Median\\$\Delta\mathrm{CI}$} & \shortstack{Rank-biserial\\$r$} & \shortstack{TOST\\$p$} & \shortstack{Holm-adj.\\$p$} & Sig. & Eq. \\
\midrule
Network & Sample & 0.00 & -0.58 & $<0.001$ & $<0.001$ & *** & Yes \\
 & Rewire & 0.00 & -0.43 & $<0.001$ & $<0.001$ & *** & Yes \\
 & Densify & 0.00 & -0.47 & $<0.001$ & $<0.001$ & *** & Yes \\
\addlinespace
Invariants & Jitter & 0.00 & -0.44 & $<0.001$ & $<0.001$ & *** & Yes \\
 & Noise & 0.00 & -0.42 & $<0.001$ & $<0.001$ & *** & Yes \\
 & Subset & 0.00 & -0.41 & $<0.001$ & $<0.001$ & *** & Yes \\
\addlinespace
Process & Bootstrapping & -0.11 & -0.82 & 0.028 & 0.028 & * & Yes \\
 & Scaling & -0.02 & -0.23 & $<0.001$ & $<0.001$ & *** & Yes \\
 & Smoothing & -0.06 & -0.73 & 0.002 & 0.003 & ** & Yes \\
\addlinespace
Signatures & Jitter & -0.01 & -0.29 & $<0.001$ & $<0.001$ & *** & Yes \\
 & Noise & -0.01 & -0.68 & $<0.001$ & $<0.001$ & *** & Yes \\
 & Subset & -0.04 & -0.82 & $<0.001$ & $<0.001$ & *** & Yes \\
\bottomrule
\end{tabularx}
\par\vspace{12pt}
\begin{minipage}{\linewidth}
\footnotesize
Frozen domain-level LOGO-CV retains original-data training fits and uses 45 paired estimator-by-domain cells (9 estimators $\times$ 5 domains), summarized over 30 seeds (42--71). EI is averaged over repetitions in both conditions. CI is computed from seed-averaged predictions for the original and deterministic conditions, but averaged over paired seed--realization metrics for stochastic perturbations. EI and CI are defined in the methodology (\nameref{sec:performance-metrics},~Eqs.~\eqref{eq:ei} and~\eqref{eq:ci}). Thus aggregation order can contribute to stochastic CI shifts. Shifts are medians of paired differences (perturbed $-$ original). Rank-biserial $r$ uses nonzero, unshifted differences with average ranks for ties; positive values favor perturbation. Only each method's most severe tested setting is included (Supplementary Table~\ref{tab:si-protocol}); Sample denotes node removal and coordinate subsetting. Original maximum targets are retained. TOST uses paired one-sided Wilcoxon signed-rank tests on $\Delta+\delta$ and $\Delta-\delta$, assuming a symmetric paired-difference distribution with location $\theta$. The nulls are $\theta\leq-\delta$ and $\theta\geq\delta$; unadjusted $p$ is the larger of the two one-sided values. Margins $\delta_{\mathrm{EI}} = 0.08$ and $\delta_{\mathrm{CI}} = 0.14$ are IQRs across the 45 original-condition cells, floored to 2 decimals. Raw differences with magnitude $<10^{-10}$ were zeroed before shifting. Tests used \texttt{SciPy 1.16.3}, \texttt{zero\_method='wilcox'}, \texttt{method='auto'}, and \texttt{correction=False}, discarding zeros after shifting. $p_{\mathrm H}$ is Holm-adjusted across the 12 methods separately for EI and CI. Eq. requires rejection of both nulls at $p_{\mathrm H}<0.05$. Shared estimators, training systems, and domains make inference conditional on this dependent panel (\nameref{sec:si-stat-conventions}).
\end{minipage}
\end{table}

\clearpage
\begin{table}[H]
\centering
\footnotesize
\captionsetup{position=top,skip=12pt}
\caption{Falsification effects under frozen and retrained frontiers.}
\label{tab:si-falsify}
\setlength{\tabcolsep}{2pt}
\begin{tabularx}{\linewidth}{@{}l>{\raggedright\arraybackslash}Xrrrrrr@{}}
\toprule
\multicolumn{8}{c}{\textbf{Domain Transfer}} \\
\cmidrule(lr){1-8}
Falsification & Method & \shortstack{Median\\$\Delta\mathrm{EI}$} & \shortstack{Rank-biserial\\$r$} & \shortstack{One-sided\\$p$} & \shortstack{Holm-adj.\\$p$} & Sig. & Diff. \\
\midrule
Frozen & Target Permutation & -0.02 & -0.38 & 0.013 & 0.013 & * & Yes \\
 & Random Generation & -0.04 & -0.91 & $<0.001$ & $<0.001$ & *** & Yes \\
 & Vector Generation & -0.03 & -0.74 & $<0.001$ & $<0.001$ & *** & Yes \\
\addlinespace
Retrain & Target Permutation & -0.09 & -0.53 & $<0.001$ & 0.001 & ** & Yes \\
 & Random Generation & -0.02 & -0.59 & $<0.001$ & $<0.001$ & *** & Yes \\
 & Vector Generation & -0.03 & -0.57 & $<0.001$ & $<0.001$ & *** & Yes \\
\midrule
\multicolumn{8}{c}{\textbf{Estimation-Observation Consensus}} \\
\cmidrule(lr){1-8}
Falsification & Method & \shortstack{Median\\$\Delta\mathrm{CI}$} & \shortstack{Rank-biserial\\$r$} & \shortstack{One-sided\\$p$} & \shortstack{Holm-adj.\\$p$} & Sig. & Diff. \\
\midrule
Frozen & Target Permutation & -0.29 & -0.97 & $<0.001$ & $<0.001$ & *** & Yes \\
 & Random Generation & -0.28 & -0.91 & $<0.001$ & $<0.001$ & *** & Yes \\
 & Vector Generation & -0.21 & -0.88 & $<0.001$ & $<0.001$ & *** & Yes \\
\addlinespace
Retrain & Target Permutation & -0.24 & -0.99 & $<0.001$ & $<0.001$ & *** & Yes \\
 & Random Generation & -0.18 & -0.94 & $<0.001$ & $<0.001$ & *** & Yes \\
 & Vector Generation & -0.31 & -0.97 & $<0.001$ & $<0.001$ & *** & Yes \\
\bottomrule
\end{tabularx}
\par\vspace{12pt}
\begin{minipage}{\linewidth}
\footnotesize
Each domain-level LOGO-CV contrast uses 45 paired estimator-by-domain cells (9 estimators $\times$ 5 domains). Predictions are averaged over 30 model-fitting seeds (42--71) before computing EI and CI within each domain. EI and CI are defined in the methodology (\nameref{sec:performance-metrics},~Eqs.~\eqref{eq:ei} and~\eqref{eq:ci}). Each falsified corpus was generated once with seed 42; model-fitting repeats are not independent negative-control corpora. Frozen frontiers retain original-data training fits, while retrained frontiers refit standardizers and estimators on the falsified training folds. Shifts are medians of paired differences (falsified $-$ original), with negative values favoring the original condition. Rank-biserial $r$ uses nonzero, unshifted differences with average ranks for ties. One-sided Wilcoxon signed-rank tests assess $H_0:\theta\geq0$ versus $H_1:\theta<0$, assuming a symmetric paired-difference distribution with location $\theta$. Raw differences with magnitude $<10^{-10}$ were zeroed. Tests used \texttt{SciPy 1.16.3}, \texttt{zero\_method='wilcox'}, \texttt{method='auto'}, and \texttt{correction=False}, discarding zero differences. $p$ is unadjusted; $p_{\mathrm H}$ is Holm-adjusted across the 6 method-by-frontier comparisons separately for EI and CI. Diff. requires $p_{\mathrm H}<0.05$ and a negative median shift, supporting inferiority to the original condition. Shared estimators, training systems, and domains make inference conditional on this dependent panel (\nameref{sec:si-stat-conventions}).
\end{minipage}
\end{table}

\clearpage
\begin{table}[H]
\centering
\footnotesize
\captionsetup{position=top,skip=12pt}
\caption{Non-inferiority of the log-additive decomposition to alternative formulations.}
\label{tab:si-ablation}
\setlength{\tabcolsep}{2pt}
\begin{tabularx}{\linewidth}{@{}l>{\raggedright\arraybackslash}Xrrrrrr@{}}
\toprule
\multicolumn{8}{c}{\textbf{Domain Transfer}} \\
\cmidrule(lr){1-8}
Ablation & Method & \shortstack{Median\\$\Delta\mathrm{EI}$} & \shortstack{Rank-biserial\\$r$} & \shortstack{One-sided\\$p$} & \shortstack{Holm-adj.\\$p$} & Sig. & NI. \\
\midrule
Simple & Invariants Only & -0.01 & -0.77 & $<0.001$ & $<0.001$ & *** & Yes \\
 & Signatures Only & -0.01 & -0.21 & $<0.001$ & $<0.001$ & *** & Yes \\
\addlinespace
Complex & Joint Features & -0.02 & -0.55 & $<0.001$ & $<0.001$ & *** & Yes \\
 & Interaction Terms & -0.01 & -0.46 & $<0.001$ & $<0.001$ & *** & Yes \\
\midrule
\multicolumn{8}{c}{\textbf{Estimation-Observation Consensus}} \\
\cmidrule(lr){1-8}
Ablation & Method & \shortstack{Median\\$\Delta\mathrm{CI}$} & \shortstack{Rank-biserial\\$r$} & \shortstack{One-sided\\$p$} & \shortstack{Holm-adj.\\$p$} & Sig. & NI. \\
\midrule
Simple & Invariants Only & -0.20 & -0.86 & $<0.001$ & $<0.001$ & *** & Yes \\
 & Signatures Only & -0.08 & -0.55 & $<0.001$ & $<0.001$ & *** & Yes \\
\addlinespace
Complex & Joint Features & -0.27 & -0.67 & $<0.001$ & $<0.001$ & *** & Yes \\
 & Interaction Terms & -0.10 & -0.65 & $<0.001$ & $<0.001$ & *** & Yes \\
\bottomrule
\end{tabularx}
\par\vspace{12pt}
\begin{minipage}{\linewidth}
\footnotesize
Each retrained domain-level LOGO-CV contrast uses 45 paired estimator-by-domain cells (9 estimators $\times$ 5 domains) and 30 model-fitting seeds (42--71), with all pipelines refit within their training folds. EI is averaged over repetitions; CI is computed from seed-averaged out-of-sample predictions within each domain. EI and CI are defined in the methodology (\nameref{sec:performance-metrics},~Eqs.~\eqref{eq:ei} and~\eqref{eq:ci}). Shifts are medians of paired differences (alternative formulation $-$ log-additive decomposition); negative values favor the decomposition. Rank-biserial $r$ uses nonzero, unshifted differences with average ranks for ties, independently of the tested margin. One-sided Wilcoxon signed-rank tests assess $H_0:\theta\geq\delta$ versus $H_1:\theta<\delta$, assuming a symmetric paired-difference distribution with location $\theta$. Margins $\delta_{\mathrm{EI}} = 0.08$ and $\delta_{\mathrm{CI}} = 0.14$ are IQRs across the 45 log-additive reference cells, floored to 2 decimals. Raw differences with magnitude $<10^{-10}$ were zeroed before subtracting $\delta$. Tests used \texttt{SciPy 1.16.3}, \texttt{zero\_method='wilcox'}, \texttt{method='auto'}, and \texttt{correction=False}, discarding zeros after shifting. $p$ is unadjusted; $p_{\mathrm H}$ is Holm-adjusted separately within the 2 simple and 2 complex alternatives for EI and CI. NI. requires $p_{\mathrm H}<0.05$ and a median shift below the margin. This supports non-inferiority within the stated margin, not superiority over the alternatives. Shared estimators, training systems, and domains make inference conditional on this dependent panel (\nameref{sec:si-stat-conventions}).
\end{minipage}
\end{table}

\clearpage
\begin{table}[H]
\centering
\footnotesize
\captionsetup{position=top,skip=12pt}
\caption{Graph invariant definitions, expressions, and asymptotic time complexity.}
\label{tab:si-features}
\begin{tabularx}{\linewidth}{@{}ll>{\raggedright\arraybackslash}Xll@{}}
\toprule
\textbf{Category} & \textbf{Invariant} & \textbf{Definition} & \textbf{Expression} & \textbf{Complexity} \\
\midrule
Simple & 1. Number of Nodes & Number of nodes\newline & $n = |V|$ & $O(1)$ \\
 & 2. Number of Edges & Number of edges\newline & $m = |E|$ & $O(1)$ \\
 & 3. Articulation Points & Vertices whose removal splits $G$ & $\lvert \{ v : \omega(G - v) > \omega(G) \} \rvert$ & $O(n+m)$ \\
 & 4. Bridges & Edges whose removal splits $G$ & $\lvert \{ e : \omega(G - e) > \omega(G) \} \rvert$ & $O(n+m)$ \\
\addlinespace
Cohesion & 5. Diameter & Longest shortest-path length \dag & $\max_v \epsilon(v)$ & $O(nm)$ \\
 & 6. Radius & Smallest vertex eccentricity \dag & $\min_v \epsilon(v)$ & $O(nm)$ \\
 & 7. Degeneracy & Largest $k$ with\newline a $k$-core & $k^{*} = \max_v c(v)$ & $O(n+m)$ \\
 & 8. $k$-Core Size & Vertices in the\newline core at $k^{*}$ & $\lvert \{ v : c(v) = k^{*} \} \rvert$ & $O(n+m)$ \\
\addlinespace
Extremal & 9. Maximum Degree & Highest vertex degree & $\max_v d_v$ & $O(n)$ \\
\addlinespace
Statistical & 10. Degree Variance & Spread of vertex\newline degrees & $n^{-1} \sum_v (d_v - \bar{d})^2$ & $O(n)$ \\
 & 11. Degree Entropy & Entropy of degree distribution & $-\sum_k p_k \ln p_k$ & $O(n)$ \\
 & 12. Joint Degree Entropy & Entropy of edge degree pairs & $-\sum_{k \le l} q_{kl} \ln q_{kl}$ & $O(m)$ \\
 & 13. Degree Skewness & Asymmetry of degree distribution & $\frac{\sqrt{n(n-1)}}{n-2}\, g_1$ & $O(n)$ \\
 & 14. Degree Kurtosis & Tail weight of degree distribution & $\frac{(n-1)[(n+1) g_2 + 6]}{(n-2)(n-3)}$ & $O(n)$ \\
 & 15. Global Clustering Coefficient & Fraction of\newline closed triples & $3 N_{\triangle} / N_{\wedge}$ & $O(n d_{\max}^2)$ \\
 & 16. Assortativity & Correlation of endpoint degrees & $\mathrm{corr}_{\{u,v\} \in E}(d_u, d_v)$ & $O(m)$ \\
\addlinespace
Spectral & 17. Normalized Laplacian 2nd Moment & Diffusion intensity & $\mathrm{tr}(L^2)/n$ & $O(m)$ \\
 & 18. Normalized Laplacian 3rd Moment & Diffusion asymmetry & $\mathrm{tr}(L^3)/n$ & $O(m^2)$ \\
 & 19. Random Walk 3rd Moment & Three-step recurrence & $\mathrm{tr}(P^3)/n$ & $O(m^2)$ \\
 & 20. Random Walk 4th Moment & Four-step recurrence & $\mathrm{tr}(P^4)/n$ & $O(m^2)$ \\
 & 21. Adjacency 4th Moment & Four-length walk density & $\mathrm{tr}(A^4)/n$ & $O(m^2)$ \\
\bottomrule
\end{tabularx}
\par\vspace{12pt}
\begin{minipage}{\linewidth}
\footnotesize
Admission required global graph descriptors defined solely by topology, independent of embeddings, weights, external parameters, and arbitrary thresholds, exactly and deterministically computable at corpus scale, and non-trivially variable across systems. All returned values were finite. \dag~Diameter and radius use the largest connected component; degeneracy and $k$-core size use the full graph. Spectral moments use exact trace identities without eigendecomposition; the 4 complete bipartite substrates use closed-form invariants. Here $n = |V|$, $m = |E|$, $d_v$ is vertex degree, $\bar{d}$ its mean, $\mu_j = n^{-1} \sum_v (d_v - \bar{d})^j$, $g_1 = \mu_3 / \mu_2^{3/2}$, and $g_2 = \mu_4 / \mu_2^{2} - 3$. $\omega$ counts connected components, $\epsilon(v)$ is eccentricity in hops, $c(v)$ is core number, $p_k$ is the fraction of nodes of degree $k$, $q_{kl}$ is the fraction of edges with unordered endpoint degrees $\{k, l\}$, and $N_{\triangle}$ and $N_{\wedge}$ count triangles and connected triples. $A$ is the adjacency matrix, $\hat{A} = D^{-1/2} A D^{-1/2}$ over non-isolated nodes, $L = I - \hat{A}$, and $P = D^{-1} A$. Isolated nodes count toward $n$ but contribute 0 to $\mathrm{tr}(L^k)$. Logarithms are natural; assortativity counts both edge orientations. Undefined assortativity and clustering are 0. Degree skewness and kurtosis are 0 for equal degrees or $n < 3$ and $n < 4$, respectively. References: \citep{ref49,ref50,ref51,ref52,ref61,ref62,ref63,ref64,ref79,ref112,ref114}.
\end{minipage}
\end{table}

\clearpage
\begin{table}[H]
\centering
\footnotesize
\captionsetup{position=top,skip=12pt}
\caption{Process signature definitions, expressions, and asymptotic time complexity.}
\label{tab:si-signatures}
\begin{tabularx}{\linewidth}{@{}ll>{\raggedright\arraybackslash}Xll@{}}
\toprule
\textbf{Property} & \textbf{Signature} & \textbf{Definition} & \textbf{Expression} & \textbf{Complexity} \\
\midrule
Dependence & 1. Lag-1 Autocorrelation & Short-term sequential dependence & $\frac{\sum_{i=1}^{L-1} \tilde{r}_i \tilde{r}_{i+1}}{\sum_{i=1}^{L} \tilde{r}_i^{2}}$ & $O(L)$ \\
 & 2. Hurst Exponent & Long-range\newline dependence & $\ln(R/s) / \ln L$ & $O(L)$ \\
\addlinespace
Dispersion & 3. Coefficient of Variation & Scale-free\newline dispersion & $s / \bar{r}$ & $O(L)$ \\
 & 4. Fano Factor & Deviation from Poisson dispersion & $s^{2} / \bar{r}$ & $O(L)$ \\
\addlinespace
Volatility & 5. Normalized Successive Difference & Local temporal volatility & $\frac{1}{(L-1)\,\bar{r}} \sum_{i=1}^{L-1} \lvert r_{i+1} - r_i \rvert$ & $O(L)$ \\
\addlinespace
Trend & 6. Trend Coefficient & Systematic non-stationarity & $\mathrm{corr}(r_i, i)$ & $O(L)$ \\
\addlinespace
Asymmetry & 7. Count Skewness & Distributional asymmetry & $\frac{1}{L} \sum_{i=1}^{L} (\tilde{r}_i / s)^{3}$ & $O(L)$ \\
\bottomrule
\end{tabularx}
\par\vspace{12pt}
\begin{minipage}{\linewidth}
\footnotesize
Admission required global series descriptors defined solely by the rate series, without external parameters or arbitrary thresholds, exactly and deterministically computable, and non-trivially variable across systems. All returned values were finite. For the chronologically ordered rate series $r_1, \ldots, r_L$ (Eq.~\eqref{eq:series}), $L = \lvert T \rvert$, $\bar{r}$ is the mean, $\tilde{r}_i = r_i - \bar{r}$, and $s$ is the sample standard deviation with denominator $L - 1$. Interval inclusion follows corpus definitions (Supplementary Table~\ref{tab:si-corpus}); lag-1 dependence and trend index consecutive included observations, not calendar gaps. The Hurst exponent is a full-series, single-window rescaled-range estimate clipped to $[0, 1]$, with $R = \max_i Y_i - \min_i Y_i$ and cumulative sums $Y_i = \sum_{j \le i} \tilde{r}_j$. Trend is Pearson correlation with position $i$. CV, Fano factor, and normalized successive difference use denominator $\bar{r} + 10^{-10}$ and are 0 when $\bar{r} \le 0$. Lag-1 autocorrelation and trend are 0 for denominators $\le 10^{-10}$. Skewness is 0 when $L < 3$ or $s \le 10^{-10}$; Hurst is 0.5 when $L < 8$ or $R$ or $s$ is $\le 10^{-10}$. References: \citep{ref53,ref54,ref55,ref65,ref66,ref108,ref109,ref110,ref113}.
\end{minipage}
\end{table}

\clearpage
\begin{table}[H]
\centering
\footnotesize
\captionsetup{position=top,skip=12pt}
\caption{Stress-test protocols and inference.}
\label{tab:si-protocol}
\begin{tabularx}{\linewidth}{@{}llXl@{}}
\toprule
\textbf{Category} & \textbf{Method} & \textbf{Description} & \textbf{Frontier} \\
\midrule
\multicolumn{4}{@{}l}{\textbf{Perturbation} (TOST equivalence at maximum intensity, $\delta_{\mathrm{EI}} = 0.08$, $\delta_{\mathrm{CI}} = 0.14$)} \\
\midrule
Network & Rewire & Degree-preserving edge swaps, fraction 0.05--0.35 (max 0.35) & Frozen \\
 & Densify & Uniform edge addition, fraction 0.05--0.35 (max 0.35) & Frozen \\
 & Sample & Uniform node removal, fraction 0.05--0.35 (max 0.35) & Frozen \\
\addlinespace
Invariants & Noise & Additive Gaussian noise, corpus-wide, 0.05--0.35 (max 0.35) & Frozen \\
 & Jitter & Multiplicative jitter, 0.05--0.35 (max 0.35) & Frozen \\
 & Subset & Random coordinate subset, retention 0.95--0.65 (max 0.65) & Frozen \\
\addlinespace
Process & Bootstrapping & Bootstrap to destroy temporal order 0.05--0.35 (max 0.05) & Frozen \\
 & Scaling & Power-law scaling, exponent 0.25--1.75 (max 0.25) & Frozen \\
 & Smoothing & Rolling-mean smoothing, window 3--15 intervals (max 15) & Frozen \\
\addlinespace
Signatures & Noise & Additive Gaussian noise, corpus-wide, 0.05--0.35 (max 0.35) & Frozen \\
 & Jitter & Multiplicative jitter, 0.05--0.35 (max 0.35) & Frozen \\
 & Subset & Random coordinate subset, retention 0.95--0.65 (max 0.65) & Frozen \\
\midrule
\multicolumn{4}{@{}l}{\textbf{Falsification} (one-sided inferiority to the original condition, Holm-adjusted)} \\
\midrule
Negative & Target Permutation & Cross-domain derangement of $y^{*}$ with $x$ and $z$ unchanged & Frozen + Retrain \\
 & Random Generation & Each feature redrawn uniformly within its observed corpus range & Frozen + Retrain \\
 & Vector Generation & Feature bootstrap plus Gaussian noise at pooled dispersion & Frozen + Retrain \\
\midrule
\multicolumn{4}{@{}l}{\textbf{Ablation} (one-sided non-inferiority of the log-additive decomposition, $\delta_{\mathrm{EI}} = 0.08$, $\delta_{\mathrm{CI}} = 0.14$)} \\
\midrule
Simple & Invariants Only & Frontier fitted on $x'$ alone & Retrain \\
 & Signatures Only & Frontier fitted on $z'$ alone & Retrain \\
\addlinespace
Complex & Joint Features & $x'$ and $z'$ concatenated into a single-stage fit & Retrain \\
 & Interaction Terms & Augmented with all 147 pairwise products & Retrain \\
\bottomrule
\end{tabularx}
\par\vspace{12pt}
\begin{minipage}{\linewidth}
\footnotesize
Stress tests use 30 model-fitting seeds (42--71), paired one-to-one with 30 independent stochastic perturbation realizations or applied to a single deterministic realization. Grids contain 7 equally spaced values between the listed endpoints; scaling includes the identity $\gamma = 1$. Feature interventions act on raw coordinates before fold-wise standardization. Additive-noise standard deviation is $s\sigma_j$, where $\sigma_j$ is the sample standard deviation of coordinate $j$ across all 25 original systems. Masking retains $\lfloor\rho p\rfloor$ coordinates, with $p = 21$ for invariants and $p = 7$ for signatures; noise, jitter draws, and masks are shared across systems within a realization. Network and process interventions recompute the affected feature vector while retaining original maximum targets. Complete bipartite node removal uses exact closed-form invariants; rewiring and densification use deterministic approximations (\hyperref[sec:si-note-bipartite]{Supplementary Note 1}), with rewiring interpolated toward configuration-model estimates \citep{ref57}. Rewiring and densification leave the complete graphs gwosc, seismic, and windmill unchanged. Falsified corpora are generated once with seed 42 as corpus-level negative controls. ``Max'' denotes the most severe setting, not necessarily the largest parameter: severity is intensity $\epsilon$ for network methods, intensity $s$ for noise and jitter, window $w$ for smoothing, $1 - \rho$ for retention $\rho$, $1 - s$ for bootstrapping, and $\lvert \ln \gamma \rvert$ for scaling exponent $\gamma$. Inference uses the 45 paired estimator-by-domain cells (9 estimators $\times$ 5 domains); metric aggregation, margins, and Holm-adjustment families are specified across test summaries (Supplementary Tables~\ref{tab:si-perturb}--\ref{tab:si-ablation}).
\end{minipage}
\end{table}

\clearpage
\subsection*{Supplementary Note 1: Complete Bipartite Perturbations}
\phantomsection
\label{sec:si-note-bipartite}
The 4 complete bipartite substrates (college, email, mooc, overflow) were represented by their fold sizes $m$ and $n$, with $N = m + n$ nodes and $M = mn$ edges, because their explicit graphs reach up to 7,200,000,000 edges. Node removal at intensity $\epsilon$ deleted $\lfloor \epsilon N \rfloor$ nodes drawn uniformly without replacement, counted the removed nodes in each fold, and evaluated the closed forms (Supplementary Table~\ref{tab:si-features}) for the remaining complete bipartite graph. This computation is exact and used 30 realizations. Rewiring and densification used a deterministic degree-class approximation with 1 realization.

For degree classes $c$ with degree $d_c$ and $N_c$ nodes, and an edge count $M'$, let
\begin{equation*}
N = \sum_c N_c, \quad S_1 = \sum_c N_c d_c, \quad S_2 = \sum_c N_c d_c^{2}, \quad \bar{d} = \frac{S_1}{N}, \quad \kappa = \frac{S_2}{\max(S_1, 1)} - 1, \quad \overline{d^{-1}} = \frac{1}{N} \sum_c \frac{N_c}{\max(d_c, 1)}.
\end{equation*}
The approximation $\Phi$ assigned the following values and set articulation points, bridges, clustering, and assortativity to 0.
\begin{itemize}[leftmargin=1.5em,itemsep=0.2em]
\item Size: $|V| = N$, $|E| = M'$, and $d_{\max} = \max_c d_c$.
\item Cohesion: diameter $D = \max\{2, \ln \max(N, 2) / \ln \kappa\}$ when $\kappa > 1$ and $D = N$ otherwise, radius $D/2$, degeneracy $k^{*} = \min(d_{\max}, \sqrt{2M'})$, and $k$-core size $\sum_c N_c \mathbf{1}[d_c \ge k^{*}]$.
\item Degree statistics: variance $N^{-1} \sum_c N_c (d_c - \bar{d})^2$, bias-corrected skewness and excess kurtosis of the class-weighted degrees, degree entropy $-\sum_k p_k \ln p_k$ over degrees rounded to integers $k$ with vertex fractions $p_k$, and joint degree entropy $-2 \sum_k q_k \ln q_k$ with stub fractions $q_k \propto k p_k$.
\item Spectral moments: $\mathrm{tr}(L^2)/N = 1 + 2M'N / \max(S_1, 1)^2$, $\mathrm{tr}(L^3)/N = 1 + (S_2 - S_1)^3 / \max(N S_1^3, 1)$, $\mathrm{tr}(P^3)/N = 6 (S_2 - S_1)^2\, \overline{d^{-1}} / [\max(6 S_1, 1) \max(S_1, 1)]$, $\mathrm{tr}(P^4)/N = N^{-1} \sum_c N_c [d_c S_2 / \max(S_1^2, 1) + \max(\kappa, 0) / \max(d_c, 1)]$, and $\mathrm{tr}(A^4)/N = [S_2 + S_2^2 / \max(2 S_1, 1)] / N$.
\end{itemize}
Rewiring at intensity $\epsilon$ evaluated $\Phi$ on the classes of $K_{m,n}$, with degrees $(n, m)$, class sizes $(m, n)$, and $M' = M$, and interpolated each invariant as $(1 - \alpha) x_j + \alpha \Phi_j$ with $\alpha = 1 - (1 - \epsilon)^2$, where $x_j$ is the exact $K_{m,n}$ value. Densification added $a = \min(\lfloor \epsilon M \rfloor, C)$ edges among the $C = \binom{m}{2} + \binom{n}{2}$ within-fold vertex pairs, so that each pair was added with probability $q = a / C$. It evaluated $\Phi$ with degrees $(n + q(m - 1), m + q(n - 1))$, class sizes $(m, n)$, and $M' = M + a$. It then multiplied the $K_{m,n}$ articulation points, bridges, and assortativity by $1 - q$ and set clustering to $\min\{c_0 + (1 - c_0) q, 1\}$, where $c_0$ is the $K_{m,n}$ clustering. All other substrates were perturbed as explicit graphs.

\clearpage
\subsection*{Supplementary Note 2: Corpus Processing Conventions}
\phantomsection
\label{sec:si-note-processing}
The \textbf{federal} series excludes awards with non-positive amounts or start dates outside 2011-01-01 to 2024-12-31. Zero speeds are treated as missing in \textbf{metrla} (8.1\% of readings) and as readings in \textbf{pemsbay} (0.003\%). \textbf{Montevideo} drops the first 4 hours removed by the loader lag and the final 20 hours of the 744-hour source that do not complete a day.

GPS timestamps in \textbf{gwosc} and \textbf{auger} were converted to UTC without leap-second correction; correction would move 19 \textbf{auger} events and no \textbf{gwosc} events to another day. The \textbf{crop} network retains 12 nodes and 30 edges from studies whose visits lack dates. Exact-name matching of \textbf{world} countries to lending metadata leaves 23 regional aggregates and spelling variants isolated.

\clearpage
\subsection*{Supplementary References}
\phantomsection
\label{sec:si-sources}
\footnotesize
\begin{enumerate}[label=S\arabic*.,ref=S\arabic*,leftmargin=2.8em,itemsep=0.6em]
\item \label{src:S1} Vallisneri, M., Kanner, J., Williams, R., Weinstein, A. \& Stephens, B. The LIGO Open Science Center. \textit{J. Phys. Conf. Ser.} \textbf{610}, 012021 (2015). \url{https://doi.org/10.1088/1742-6596/610/1/012021}
\item \label{src:S2} U.S. Geological Survey. USGS Water Data for the Nation (1994). \url{https://doi.org/10.5066/F7P55KJN}
\item \label{src:S3} Pierre Auger Collaboration. Pierre Auger Observatory Open Data, version 3 [Dataset]. Zenodo (2024). \url{https://doi.org/10.5281/zenodo.10488964}
\item \label{src:S4} Albuquerque Seismological Laboratory/USGS. Global Seismograph Network (GSN - IRIS/USGS) [Dataset]. International Federation of Digital Seismograph Networks (1988). \url{https://doi.org/10.7914/SN/IU}
\item \label{src:S5} U.S. Geological Survey. Advanced National Seismic System (ANSS) Comprehensive Catalog (2017). \url{https://doi.org/10.5066/F7MS3QZH}
\item \label{src:S6} Meteostat. Meteostat weather and climate database. \url{https://meteostat.net} (accessed 1 September 2026).
\item \label{src:S7} Cassada, R. C. \& Russell, R. L. The dauerlarva, a post-embryonic developmental variant of the nematode \textit{Caenorhabditis elegans}. \textit{Dev. Biol.} \textbf{46}, 326--342 (1975). \url{https://doi.org/10.1016/0012-1606(75)90109-8}
\item \label{src:S8} Golden, J. W. \& Riddle, D. L. The \textit{Caenorhabditis elegans} dauer larva: developmental effects of pheromone, food, and temperature. \textit{Dev. Biol.} \textbf{102}, 368--378 (1984). \url{https://doi.org/10.1016/0012-1606(84)90201-X}
\item \label{src:S9} Allen-Perkins, A. et al. CropPol: A dynamic, open and global database on crop pollination. \textit{Ecology} \textbf{103}, e3614 (2022). \url{https://doi.org/10.1002/ecy.3614}
\item \label{src:S10} Hungarian Chickenpox Cases [Dataset]. UCI Machine Learning Repository (2021). \url{https://doi.org/10.24432/C5103B}
\item \label{src:S11} U.S. Food and Drug Administration. openFDA Drug Event API. \url{https://open.fda.gov/apis/drug/event/} (accessed 13 March 2026).
\item \label{src:S12} Guttag, J. CHB-MIT Scalp EEG Database, version 1.0.0 [Dataset]. PhysioNet (2010). \url{https://doi.org/10.13026/C2K01R}
\item \label{src:S13} U.S. Department of the Treasury. USAspending API. \url{https://api.usaspending.gov/} (accessed 25 August 2026).
\item \label{src:S14} Kumar, S., Zhang, X. \& Leskovec, J. Predicting dynamic embedding trajectory in temporal interaction networks. In \textit{Proc. 25th ACM SIGKDD Int. Conf. Knowledge Discovery and Data Mining} 1269--1278 (ACM, 2019). \url{https://doi.org/10.1145/3292500.3330895}
\item \label{src:S15} McAuley, J., Targett, C., Shi, Q. \& van den Hengel, A. Image-based recommendations on styles and substitutes. In \textit{Proc. 38th Int. ACM SIGIR Conf. Research and Development in Information Retrieval} 43--52 (ACM, 2015). \url{https://doi.org/10.1145/2766462.2767755}
\item \label{src:S16} Kumar, S., Spezzano, F., Subrahmanian, V. S. \& Faloutsos, C. Edge weight prediction in weighted signed networks. In \textit{Proc. IEEE 16th Int. Conf. Data Mining} 221--230 (IEEE, 2016). \url{https://doi.org/10.1109/ICDM.2016.0033}
\item \label{src:S17} World Bank. Projects \& Operations API and Country API. \url{https://search.worldbank.org/api/v2/projects} and \url{https://api.worldbank.org/v2/country} (accessed 29 September 2026).
\item \label{src:S18} Wikipedia Math Essentials [Dataset]. UCI Machine Learning Repository (2021). \url{https://doi.org/10.24432/C5V32F}
\item \label{src:S19} Paranjape, A., Benson, A. R. \& Leskovec, J. Motifs in temporal networks. In \textit{Proc. 10th ACM Int. Conf. Web Search and Data Mining} 601--610 (ACM, 2017). \url{https://doi.org/10.1145/3018661.3018731}
\item \label{src:S20} Panzarasa, P., Opsahl, T. \& Carley, K. M. Patterns and dynamics of users' behavior and interaction: network analysis of an online community. \textit{J. Am. Soc. Inf. Sci. Technol.} \textbf{60}, 911--932 (2009). \url{https://doi.org/10.1002/asi.21015}
\item \label{src:S21} OpenStreetMap contributors. OpenStreetMap [Dataset]. \url{https://www.openstreetmap.org} (accessed 10 November 2025).
\item \label{src:S22} Boeing, G. Modeling and analyzing urban networks and amenities with OSMnx. \textit{Geogr. Anal.} \textbf{57}, 567--577 (2025). \url{https://doi.org/10.1111/gean.70009}
\item \label{src:S23} Li, Y., Yu, R., Shahabi, C. \& Liu, Y. Diffusion convolutional recurrent neural network: data-driven traffic forecasting. In \textit{Int. Conf. Learning Representations} (2018).
\item \label{src:S24} Abbott, B. P. et al. GWTC-1: A Gravitational-Wave Transient Catalog of Compact Binary Mergers Observed by LIGO and Virgo during the First and Second Observing Runs. \textit{Phys. Rev. X} \textbf{9}, 031040 (2019). \url{https://doi.org/10.1103/PhysRevX.9.031040}
\item \label{src:S25} Abbott, R. et al. GWTC-2: Compact Binary Coalescences Observed by LIGO and Virgo during the First Half of the Third Observing Run. \textit{Phys. Rev. X} \textbf{11}, 021053 (2021). \url{https://doi.org/10.1103/PhysRevX.11.021053}
\item \label{src:S26} Abbott, R. et al. GWTC-2.1: Deep extended catalog of compact binary coalescences observed by LIGO and Virgo during the first half of the third observing run. \textit{Phys. Rev. D} \textbf{109}, 022001 (2024). \url{https://doi.org/10.1103/PhysRevD.109.022001}
\item \label{src:S27} Abbott, R. et al. GWTC-3: Compact Binary Coalescences Observed by LIGO and Virgo during the Second Part of the Third Observing Run. \textit{Phys. Rev. X} \textbf{13}, 041039 (2023). \url{https://doi.org/10.1103/PhysRevX.13.041039}
\item \label{src:S28} Government of Uruguay. National Open Data Catalog. \url{https://catalogodatos.gub.uy/} (accessed 1 September 2026).
\item \label{src:S29} Kunz, N. \& Gao, H. O. Global geolocated realtime data of interfleet urban transit bus idling. Preprint at \url{https://doi.org/10.48550/arXiv.2403.03489} (2024).
\item \label{src:S30} LIGO Scientific Collaboration, Virgo Collaboration \& KAGRA Collaboration. GWTC-4.0 Data Release Documentation [Dataset]. Gravitational Wave Open Science Center (2025). \url{https://doi.org/10.7935/aes8-px89}
\item \label{src:S31} Rader, R. et al. Non-bee insects are important contributors to global crop pollination. \textit{Proc. Natl Acad. Sci. USA} \textbf{113}, 146--151 (2016). \url{https://doi.org/10.1073/pnas.1517092112}
\end{enumerate}

\endgroup
\end{document}